\documentclass[aps,prr,superscriptaddress,twocolumn%,longbibliography
]{revtex4-1}
\usepackage{graphicx,amsmath,mathtools,amsfonts}
\usepackage{bm}
\usepackage[normalem]{ulem}
\usepackage[usenames, dvipsnames]{xcolor}
\usepackage{multirow}
\usepackage{makecell}
\usepackage[utf8]{inputenc}

\usepackage{hyperref}
\hypersetup{colorlinks=true, linkcolor=blue, citecolor=blue, urlcolor=blue}

\begin{document}

\title{Ruddlesden-Popper and perovskite phases as a material platform for altermagnetism}

\author{F. Bernardini}
\affiliation{Dipartimento di Fisica, Universit\`a di Cagliari, IT-09042 Monserrato, Italy}
\author{M. Fiebig}
\affiliation{Department of Materials, ETH Zurich, Vladimir-Prelog-Weg 4, 8093 Zurich, Switzerland}
\author{A. Cano}
\affiliation{
Univ. Grenoble Alpes, CNRS, Grenoble INP, Institut Néel, 25 Rue des Martyrs, 38042, Grenoble, France
}
\date{\today}

\begin{abstract}
The subclass of collinear antiferromagnets where spin Kramers degeneracy is broken---resulting in ferromagnetic-like properties---offers exciting new opportunities in magnetism and hence motivates the reasoned expansion of the material base for these so-called altermagnets. Here, we show that Ruddlesden-Popper and perovskite phases are generic hosts for altermagnetic behavior. Using first-principles calculations, we demonstrate altermagnetism in prototypical nickel-based systems such as La$_2$NiO$_4$ and identify additional candidates, including the superconducting La$_3$Ni$_3$O$_7$ and the multiferroic BiFeO$_3$. 
These materials span insulating, semiconducting, and metallic conduction types, with computed nonrelativistic spin splittings reaching up to 250~meV. Our analysis also reveals the presence of accidental nodes and distinct topologies in the spin-momentum texture at the Brillouin-zone boundary, suggesting a refined classification for altermagnetic materials beyond the $d$-wave or higher even-parity wave classes. In addition, we address formal inconsistencies in the traditional classification of magnetically ordered systems, proposing resolutions grounded in the altermagnetic framework, and point out the potential for altermagnetic behavior in systems beyond collinear antiferromagnets with perfectly compensated magnetization, broadening the scope for future exploration.
\end{abstract}

\maketitle

\section{Introduction}

Ferromagnets and collinear antiferromagnets have long been believed to display fundamentally distinct properties due to the presence vs absence of Kramers spin degeneracy---that is, the presence/absence of (nonrelativistic) spin splitting of their energy bands. 
However, there is an increasing body of both theoretical and experimental evidence showing that there exists a subclass of collinear antiferromagnets lacking such a degeneracy \cite{kusunose19,ahn19,Naka2019,smejkal20,mokrousov20,zunger20-prb,Feng2022,CUONO2023171163,zunger23,GUO2023100991,ang23a,
ang23b,
ang23c,
gao23a,gao23b,gao24,fernandes24-prb,sato24-prb}.
These systems, termed altermagnets \cite{jungwirth22-prx,mazin22-prx}, offer new exciting perspectives in magnetism by combining the best of two worlds \cite{manchon22-jpd:ap,yao24-review}. 
On one hand, their spin splitting enables them to display ferromagnetic(FM)-like responses that can be useful for spintronic applications.
These include anomalous Hall effect, giant or tunneling magnetoresistance, spin-polarized currents, and magnetooptics. 
On the other hand, the precession frequencies associated with antiferromagnetic (AFM) order can be higher and the zero net magnetization obtained in this case eliminates undesired stray magnetic fields. 
These are a critical requirements for miniaturized spintronic technologies with faster devices. 

\begin{figure}[b!]
\includegraphics[width=.46\textwidth]{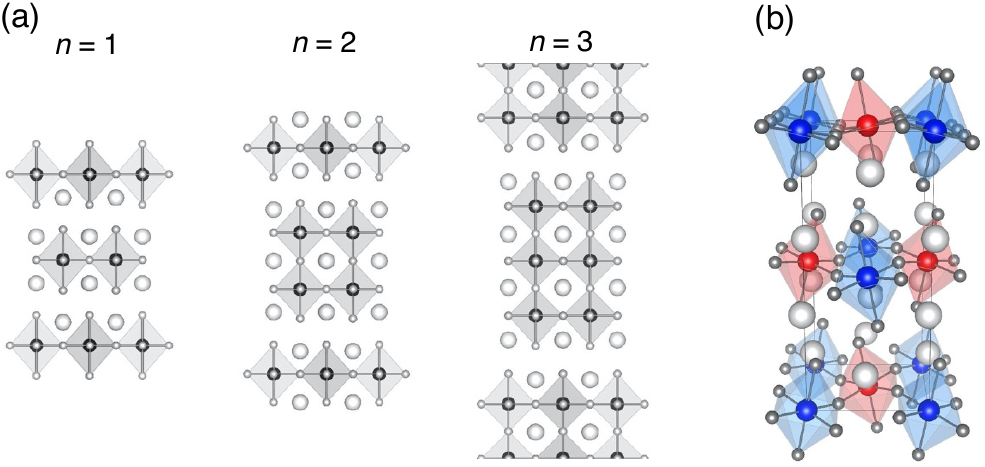}
    \caption{(a) Ideal structure of the $n =1,2,$ and 3 members of the Ruddlesden-Popper series and (b) example of the structural distortions that generally take place in these systems. 
    The red/blue balls and octahedra in (b) indicate opposite spin-reversed sublattices associated with $G$-type AFM order. 
    }
    \label{structure}
\end{figure}

Here, we introduce the rich series of Ruddlesden-Popper phases as a versatile playground for altermagnetism, including their perovskite end members.
These phases have the general chemical formula $A_{n-1}A’_2B_nX_{3n+1}$ ($ABX_3$ in the $n=\infty$ limit), where $A$ and $A'$ are alkali, alkaline-earth, or rare-earth metals, $B$ is a transition metal, and $X $ is an anion such as O or F, for example.
The ideal crystal structure of these materials is illustrated in Fig. \ref{structure} (a). 
It displays $n$ perovskite-like layers in which the $B$ atoms are surrounded by $X$ octahedra, 
which are further sandwiched between two $A'X$ layers that can be regarded as rock-salt-type spacers. 
Similarly to the parent perovskites, the Ruddlesden-Popper phases are prone to structural distortions involving the tilting of the anion octahedra such as the one illustrated in Fig. \ref{structure} (b). 
Consequently, the emergence of AFM order in crystal setups of this type may readily lack the combined time-reversal and translation or space-inversion symmetries necessary to enforce Kramers spin degeneracy.

We chose prototypical nickelates such as La$_2$NiO$_4$ to illustrate the potential of the Ruddledsen-Popper platform in terms of AM materials. 
Besides spintronic applications, our choice is motivated by the distinct fundamental interplay that can be expected between altermagnetism and unconventional superconductivity \cite{mazin22}, as the latter has recently been reported in these systems 
\cite{wang23-bilayer,wen23-trilayer,mundy21,review-nickelates20}. 
In addition, to illustrate the large variety of AM materials that can be found within the Ruddledsen-Popper class, we demonstrate the AM character of some original mixed-anion variants ($X =$~O/F) and identify additional materials with other transition-metal atoms hosting AM properties
($B =$~Co, Fe, Mn, and Cr). 
The latter includes perovskites such as the prototypical multiferroic material BiFeO$_3$, which we propose as a test-bed material in relation to the defining properties of altermagnetism.

\section{Computational methods}

We performed density-functional-theory calculations using the all-electron code {\sc{WIEN2k}} based on the full-potential augmented plane-wave plus local-orbitals method (APW+LO)~\cite{WIEN2k}. 
We used the structural parameters determined experimentally and either the local density approximation (LDA) or the generalized gradient approximation in its Perdew-Burke-Ernzerhof (PBE) form for the exchange-correlation functional \cite{LDA,PBE}. 
The precise choice will be done to best reproduce the experimental magnetic moments (or according to previous theory if no experimental value has been reported yet). 
We used muffin-tin radii of 2.5, 2.0, and 1.60 a.u. for the La, Ni, and O (F) atoms, respectively, and a plane-wave cutoff $R_{\rm MT}K_{\rm max}$ = 6.0. 
The integration over the Brillouin zone was done using adapted Monkhorst-Pack meshes with a $k$-point distance $\leq$~0.1{\AA$^{-1}$} and denser meshes for the Fermi and isoenergy surfaces.

\section{Altermagnetic nickelates}

\subsection{Single-layer %($n = 1$) 
AM nickelates}

We first consider the representative single-layer nickelate La$_2$NiO$_4$ ($n = 1$ in the Ruddlesden-Popper series). 
At room temperature, the crystal structure of this system corresponds to the orthorhombic $Bmab$ space group with tilted oxygen octahedra \footnote{$Bmab$ is a nonconventional setting of $Cmca$ where the $c$ axis corresponds to the long (stacking) axis. In the new ITA notation, they are denoted as $Bmeb$ and $Cmce$ respectively.}. 
The Ni atoms are at the 4$b$ Wyckoff positions and display a $G$-type AFM order as can be described within the crystallographic unit cell as illustrated in Fig. \ref{structure}(b) \cite{rodriguez-carvajal91}. As a result, the opposite-spin sublattices are connected by rotation, but not by translation or inversion.  
We note that La$_2$NiO$_4$ is isostructural and displays a similar AFM order than the cuprate superconductor La$_2$CuO$_4$, which has been previously pointed out as AM \cite{jungwirth22-prx}. 
The $G$-AFM order in La$_2$NiO$_4$, however, is different in the sense that the spins point along the $x$ direction. 
This may be relevant in relation to relativistic effects (which we neglect hereafter).  

\begin{figure}[t!]
\includegraphics[width=.475\textwidth]{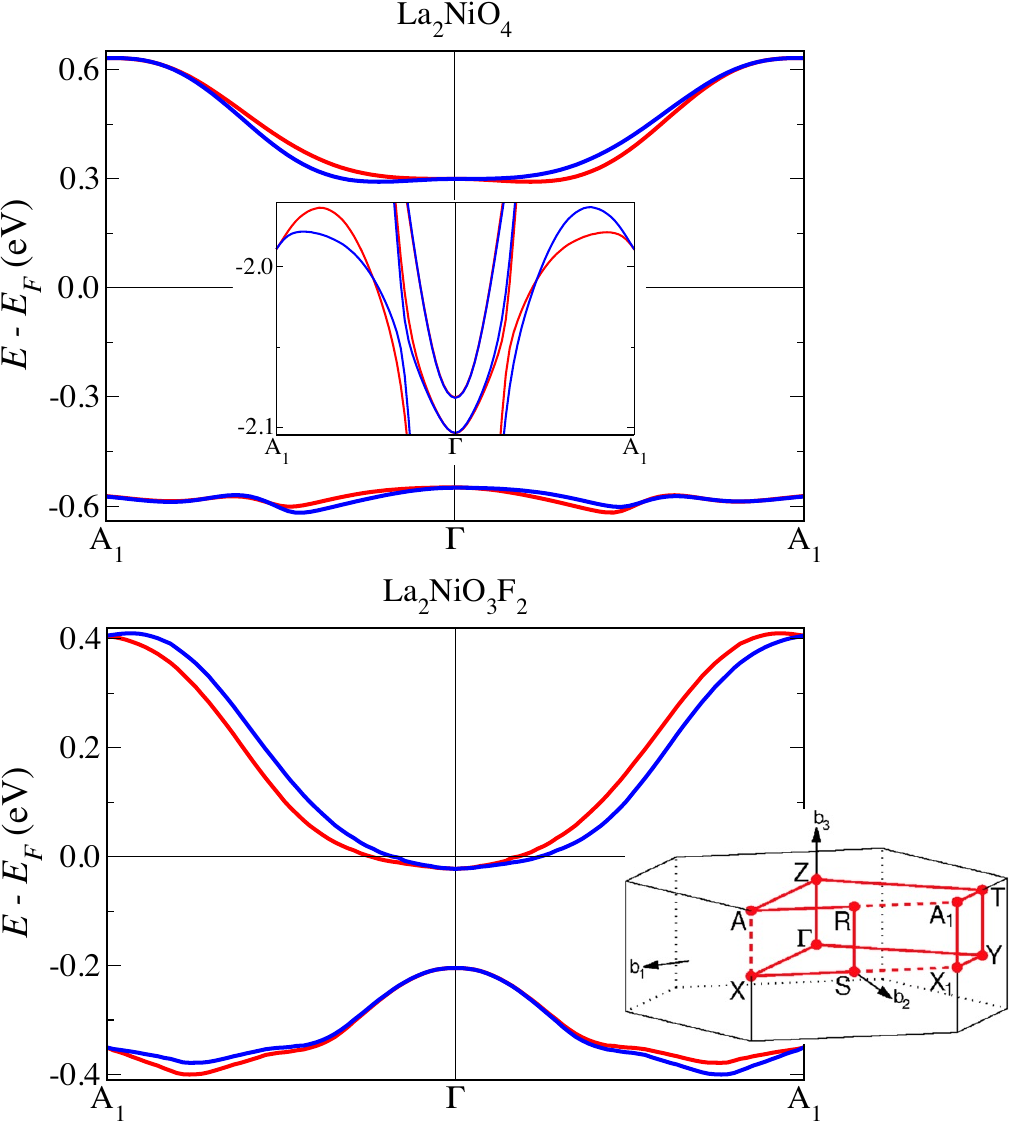}
    \caption{
    Spin-projected band structure of the single-layer nickelates La$_2$NiO$_4$ and La$_2$NiO$_3$F$_2$
    in their collinear $G$-AFM state [see Fig. \ref{structure}(b)]. 
    The splitting of the `up' and `down' (red and blue) bands is a manifestation of altermagnetism (no spin-orbit coupling is considered). In the top panel, the inset illustrates the presence of accidental nodes where spin degeneracy is restored. In the bottom panel, the inset (from \cite{Setyawan_2010}) illustrates the Brillouin zone of these systems and its high-symmetry points (the A$_1$ points in the band plots correspond to different equivalent points).
    }
    \label{single-layer}
\end{figure}

Figure \ref{single-layer} illustrates the broken spin degeneracy that results from the collinear $G$-AFM order in La$_2$NiO$_4$. 
The spin splitting of the bands can be expected to scale with the Ni magnetic moment. Experimentally, this is $1.05$~$\mu_{\rm B}$ at room temperature (and saturates to $1.68$~$\mu_{\rm B}$ at 4~K). Using the LDA for the exchange-correlation functional, we find 1.1~$\mu_{\rm B}$, while using PBE, we obtain 1.3~$\mu_{\rm B}$. However, the maximum splitting of the valence bands is $\sim 161$~meV in both cases as summarized in Table \ref{tab:summary}.

\begin{table*}[t!]
    \centering
    \begin{tabular}{l c c c c c c c c}
\hline \hline         
 && \multirow{2}{5em}{Conduction} && \multirow{2}{9em}{Spin splitting (meV)}  && \raisebox{-.5em}{\makecell[cc]{Spin projected \\ Fermi/isoenergy surface}} && \multirow{2}{4em}{$T_{\rm AM}$ (K)} \\
 \hline
 La$_2$NiO$_4$ (single-layer, $n =1$) && I && - $\mid$ - $\mid$ 161 ($-$4.42)  && \raisebox{-.5\height}{\includegraphics[height=.125\textwidth]{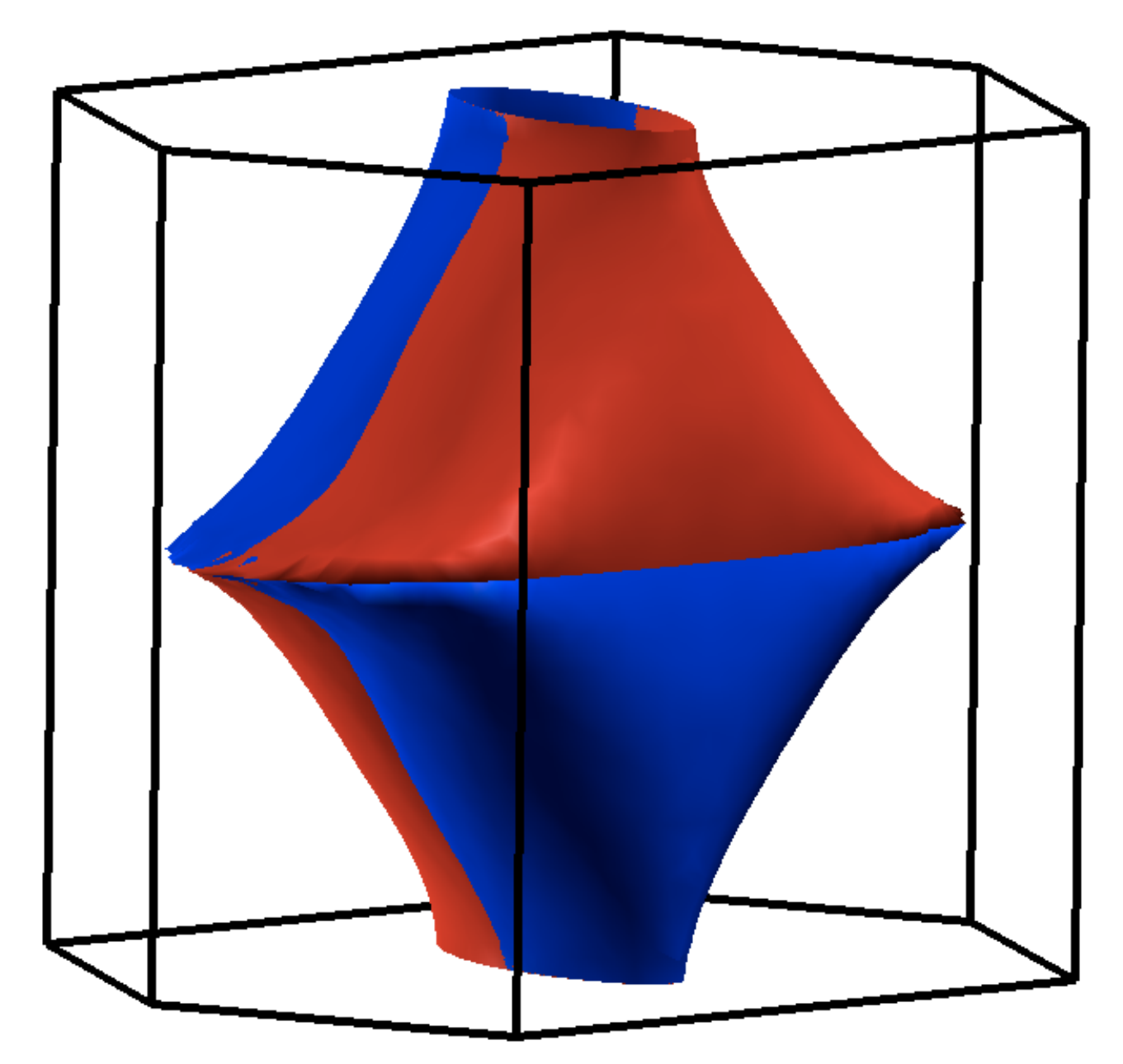}} 
 %& \raisebox{-.5\height}{\includegraphics[height=.1\textwidth]{fig12.eps}}
&& 330 \\
La$_2$NiO$_3$F$_2$ (single-layer, $n =1$) && M && 86 $\mid$ 112 $\mid$ 194 ($-$1.49)   && \raisebox{-.5\height}{\includegraphics[height=.125\textwidth]{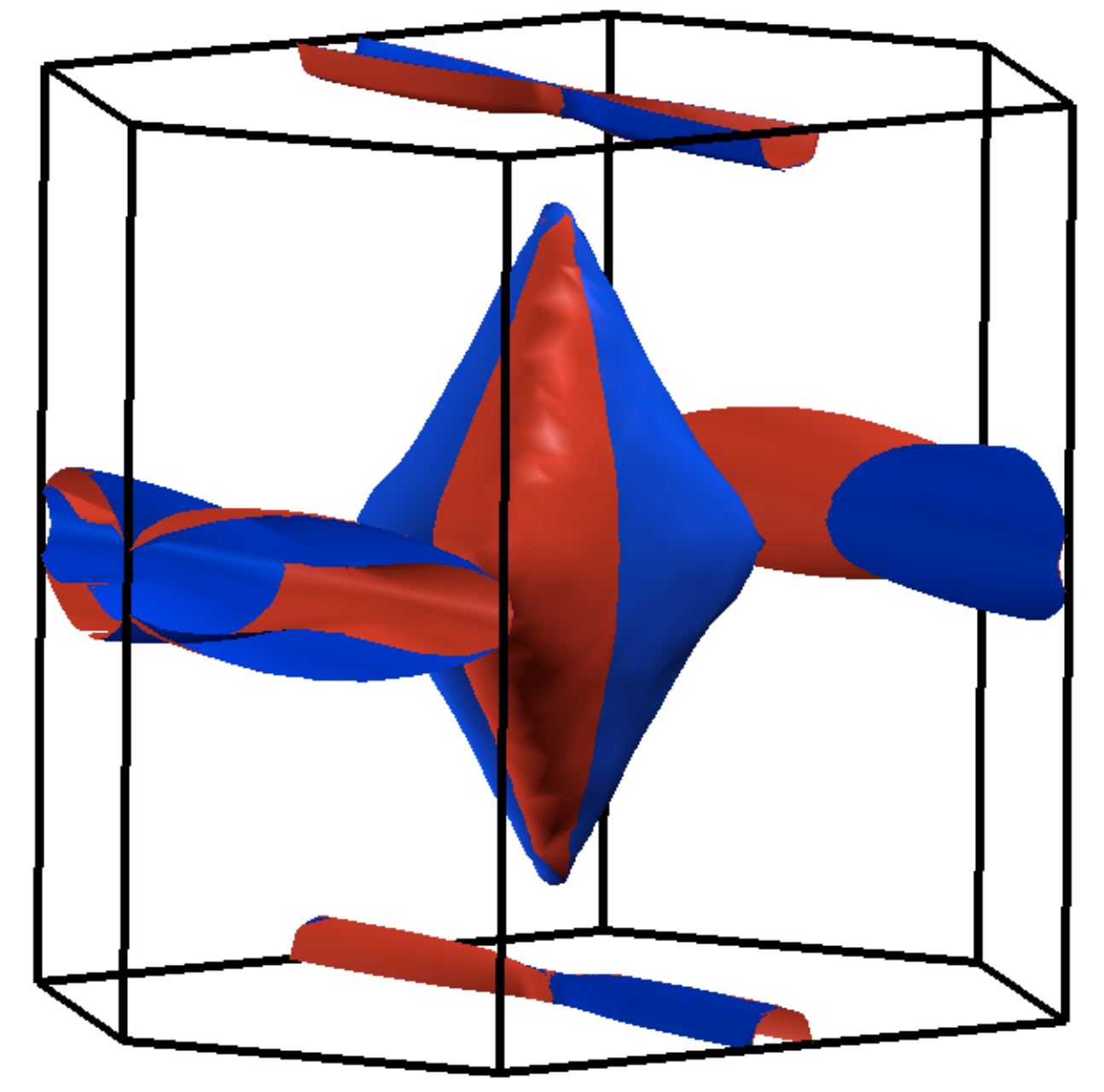}}
&& $>55$\\
La$_2$NiO$_3$F$_{(2-1/16)}$ (single-layer, $n =1$) && M &&77 $\mid$ 132 $\mid$ 251 ($-$0.76)   && \raisebox{-.5\height}{\includegraphics[height=.125\textwidth]{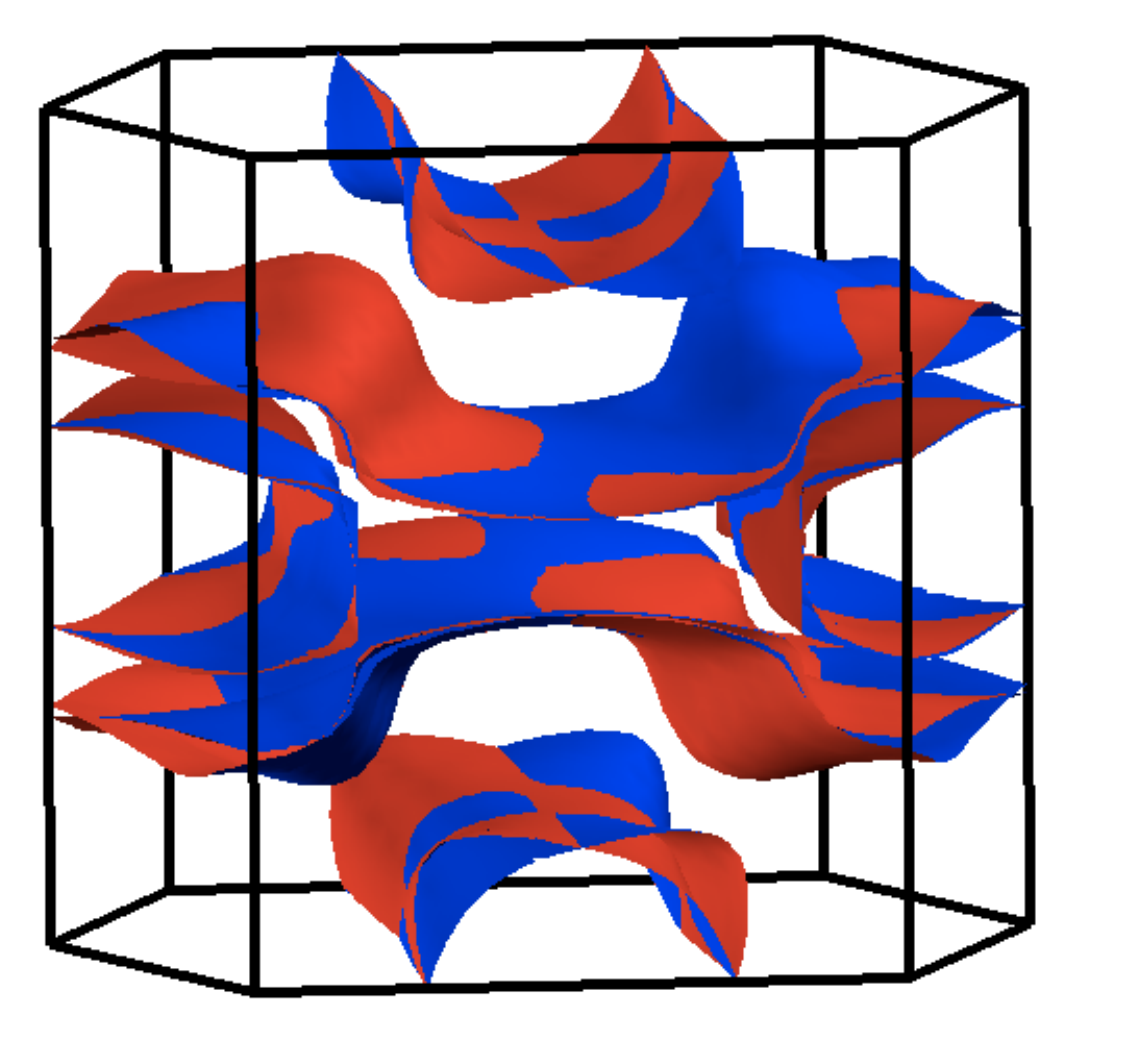}} 
%& \raisebox{-.5\height}{\includegraphics[height=.125\textwidth]{fig20.eps}}
&& $>100$ \\
%\hline
La$_3$Ni$_2$O$_7$ (bilayer, $n =2$) && M && 28 $\mid$ 48 $\mid$ 121 ($-$3.92)  && 
\raisebox{-.5\height}{\includegraphics[height=.125\textwidth]{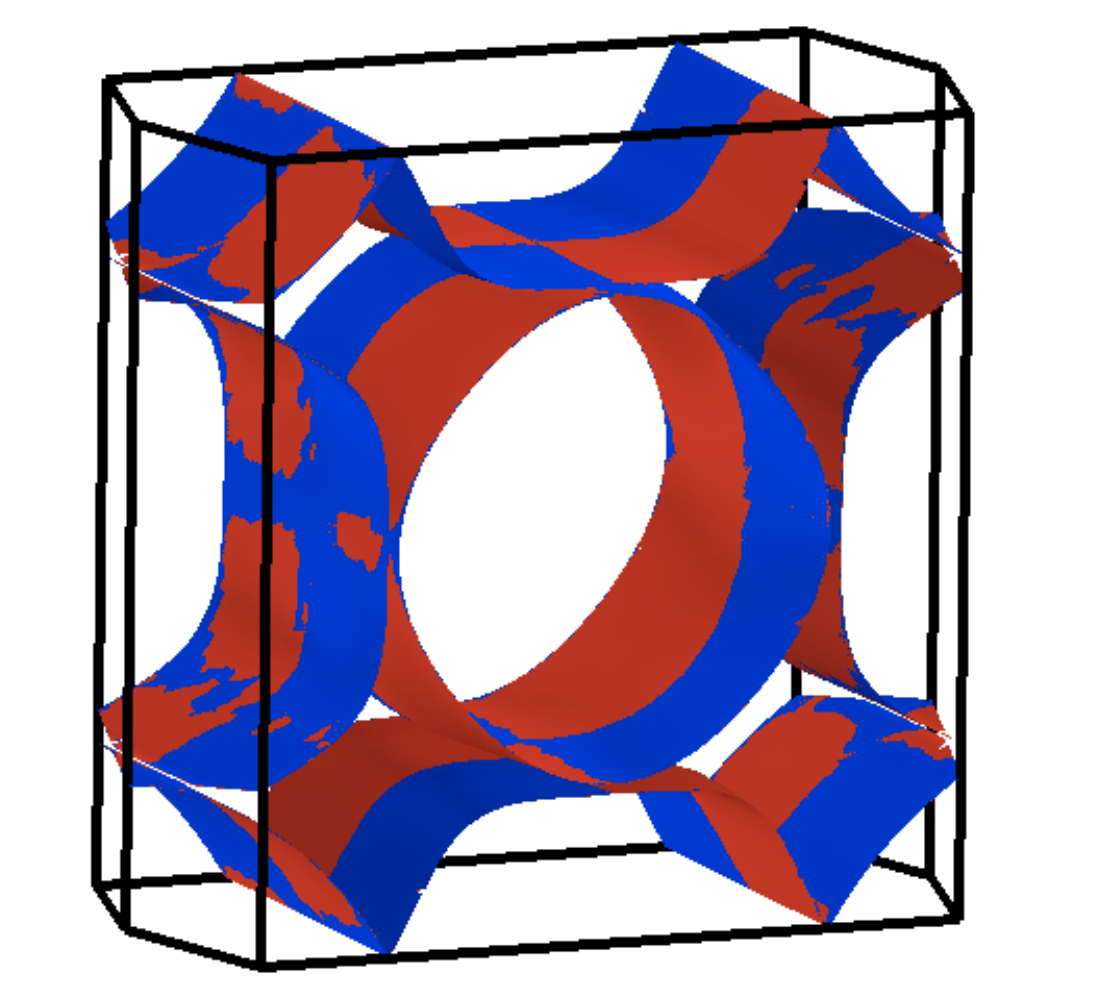}}
&& $\sim 153$ \\
LaNiO$_2$ (infinite-layer, $n =\infty$) && M  && 9 $\mid$ 25 $\mid$ 170 ($-$1.77)   && 
\raisebox{-.5\height}{\includegraphics[height=.125\textwidth]{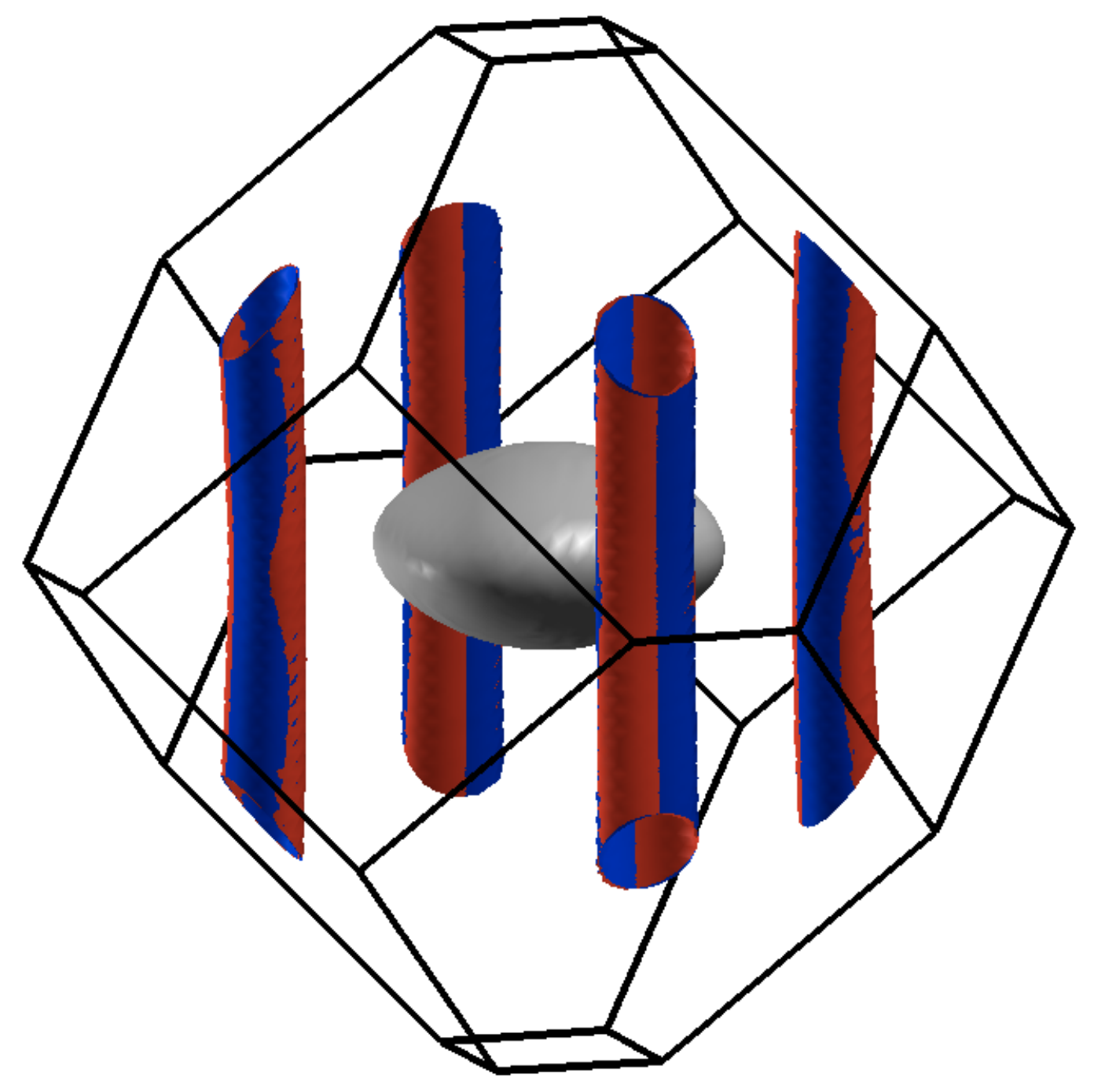}} 
&& -\\
%\hline
BiNiO$_3$ (perovskite, $n =\infty$) && \makecell[cc]{I ($P\bar{1}$) \\ M ($Pbnm$)} && 126 $\mid$ 149 $\mid$ 167 ($-$1.48)   &&
\raisebox{-.5\height}{\includegraphics[height=.125\textwidth]{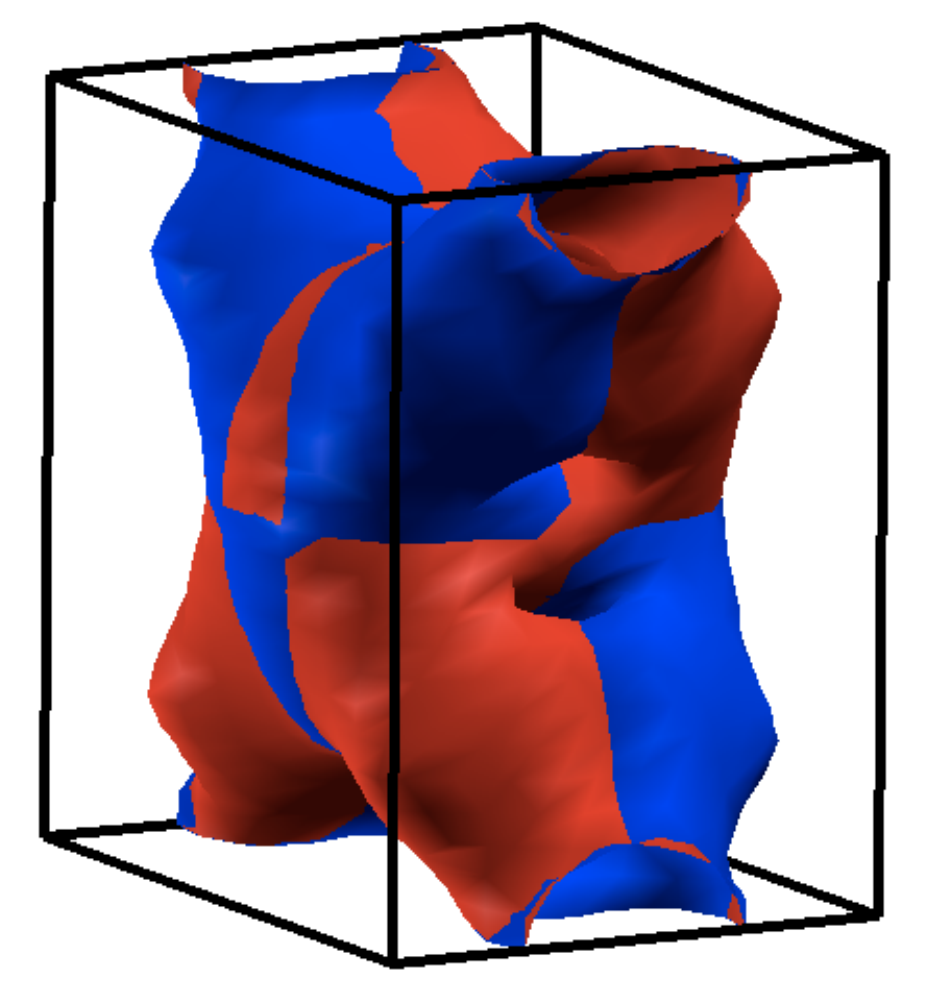}}
 && 300 \\
PbNiO$_3$ (perovskite, $n =\infty$) && I && - $\mid$ - $\mid$ 334 ($-$4.05)  &&
\raisebox{-.5\height}{\includegraphics[height=.125\textwidth]{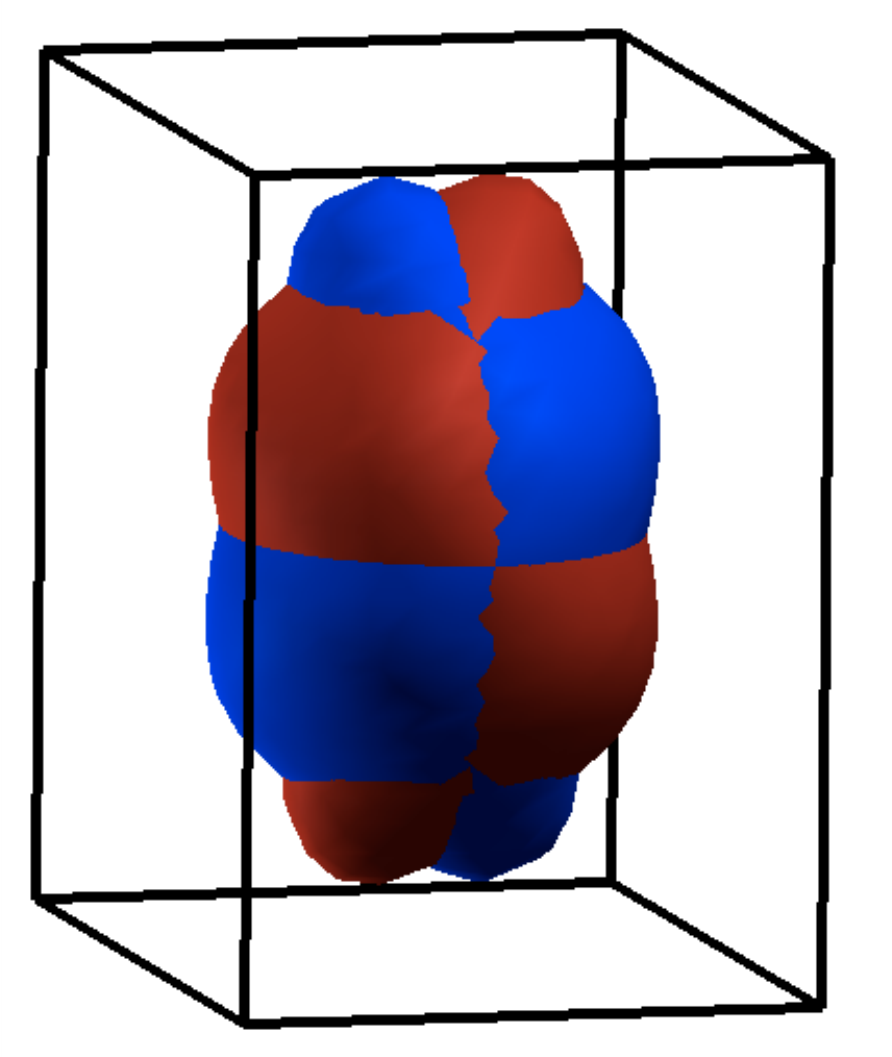}}
&& \makecell[cc]{225 ($Pbnm$) \\ 205 ($R3c$)} \\
BiFeO$_3$ (perovskite, $n =\infty$) && I && - $\mid$ - $\mid$ 316 ($-$5.53)   &&
\raisebox{-.5\height}{\includegraphics[height=.125\textwidth]{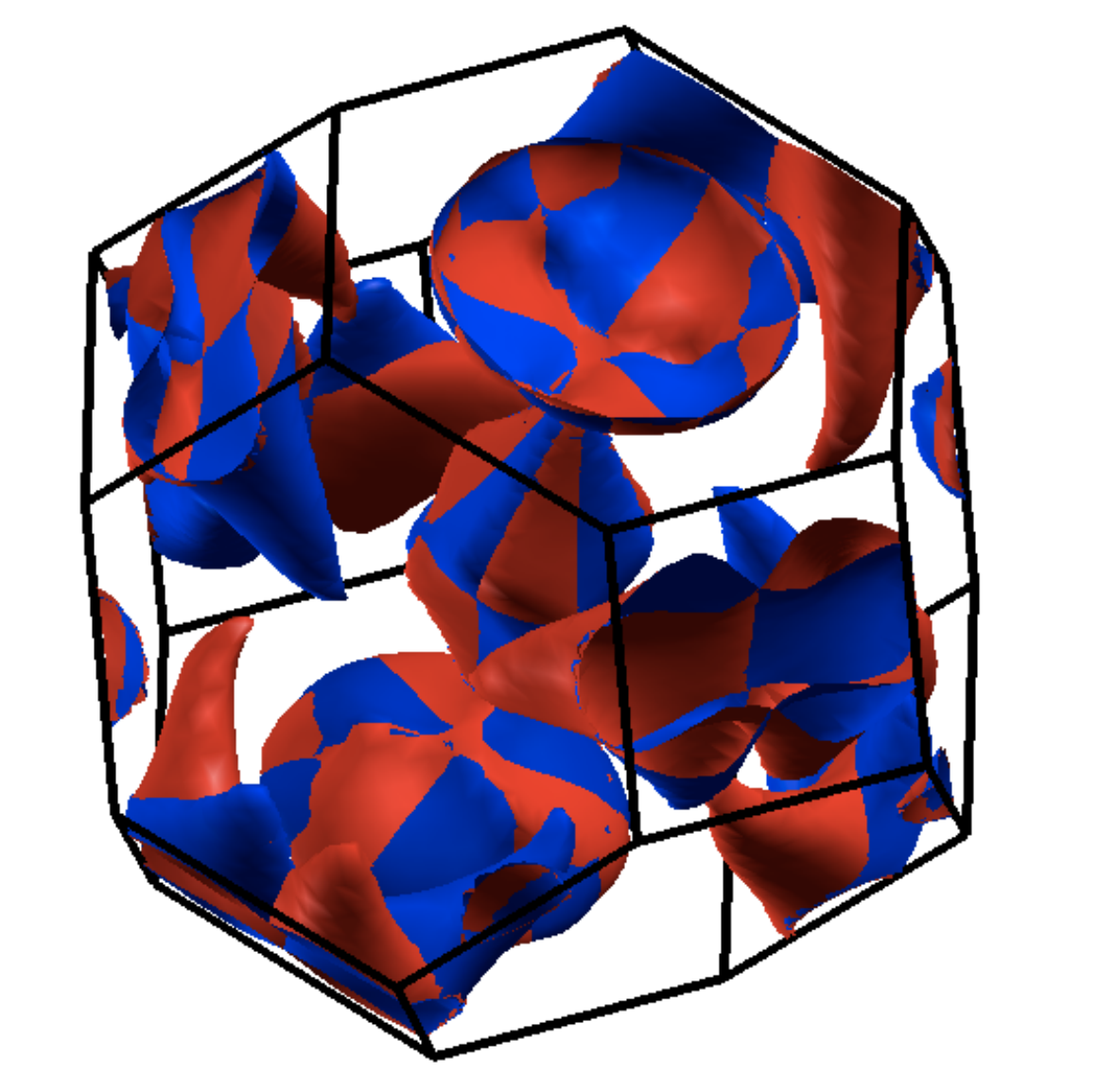}}
&& 643 \\
\hline 
\multicolumn{6}{l}{Additional AM materials}\\
\multicolumn{6}{l}{(Sr,La)$_2$MnO$_4$, La$_2$CoO$_4$, Ca$R$CrO$_4$ ($R =$~Pr, Nd, Sm, and Eu)}
\\
\hline \hline         
    \end{tabular}
    \caption{
    Summary of the main AM features of the selected materials within the Ruddlesden-Popper and perovskite series. M and I indicate metallic and insulating conduction respectively according to DFT calculations, while the spin splitting refers to the maximal splitting at the Fermi level (first value, only defined in metals), the maximal splitting of the bands crossing the Fermi level (second value, only defined in metals), and the maximal splitting of the valence bands with the energy relative to the Fermi level at which it happens between brackets (third value). $T_{\rm AM}$ indicates the magnetic transition temperature at which AM properties emerge ({\it i.~e.} the Néel temperature). 
    }
    \label{tab:summary}
\end{table*}

Next, we consider the isoelectronic mixed-anion compound La$_2$NiO$_3$F$_2$, which displays the same type of $G$-AFM order with the spins pointing along the small crystal axis \cite{clemens18,clemens20}. 
The crystal structure, however, is different since the anion-octahedra tiltings are in antiphase along the out-of-plane $c$ direction. Also, there is an ordering of F and O atoms at apical and interstitial sites respectively. As a result, the crystal symmetry corresponds to the orthorhombic $Cccm$ space group with the Ni atoms at the 4$e$ Wyckoff positions. Yet, La$_2$NiO$_3$F$_2$ should also be AM since the opposite-spin sublattices are connected by rotation only.  

Figure \ref{single-layer} shows the computed band structure for La$_2$NiO$_3$F$_2$. The splitting of the bands confirms the AM character of this system, which interestingly we find to be metallic. Experimentally, the Ni magnetic moment is $0.7$~$\mu_{\rm B}$ at 10~K \cite{clemens20}. 
Using LDA we find $1.08$~$\mu_{\rm B}$, which further gives a maximum spin splitting of 86~meV at the Fermi level. This splitting reaches 112~meV for the bands that cross the Fermi level and 194~meV for the valence bands. These values are summarized in Table \ref{tab:summary}.

The mixed-anion strategy can be exploited to optimize altermagnetism and metallicity by reducing the previous single-layer nickelates towards La$_2$NiO$_3$F, which is a candidate route for promoting superconductivity \cite{cano21-prm,cano22-chemmat}. 
The intermediate compound La$_2$NiO$_3$F$_{(2 - 1/16)}$, in particular, has been synthesized in \cite{clemens20}. 
The crystal structure of this system corresponds to the space group $C12/c1$ where the Ni atoms are at the 4$c$ Wyckoff positions. The tilting of the anion octahedra is therefore supplemented with a monoclinic distortion in this case. Compared to La$_2$NiO$_3$F$_2$, this system displays the same type of $G$-AFM order with a magnetic moment of $1.62$~$\mu_{\rm B}$ at 10~K, and the order is retained up to higher temperatures. 
La$_2$NiO$_3$F$_{(2 - 1/16)}$ thus emerges as an additional metallic altermagnet whose spin splitting is $\gtrsim$ 77 meV (see Table \ref{tab:summary}).

\subsubsection{Trivial and non-trivial topologies of the spin-momentum texture at the boundary of the Brillouin zone}

We note that, even if the AFM order in La$_2$NiO$_4$ and La$_2$NiO$_3$F$_2$ is the same, their different crystal structure makes them qualitatively different in terms of altermagnetism. 
The Brillouin zone of AM materials is always divided into an even number of sectors where the spin splitting is pair-wise reversed (spin compensation). This implies the presence of nodes at which the spin degeneracy is restored. These symmetry-imposed nodes have previously been discussed in analogy with unconventional superconductivity in terms of $d$-wave (or higher even-parity wave) symmetry \cite{jungwirth22-prx}.
For La$_2$NiO$_4$ and La$_2$NiO$_3$F$_2$, these nodes are illustrated by the spin-projected isoenergy surfaces shown in Table \ref{tab:summary}. As we see, these systems display two perpendicular nodal planes associated with high-symmetry planes of the Brillouin zone.  
One of them ($k_y = 0$) is common to both these systems. 
In La$_2$NiO$_4$, the other plane corresponds to $k_z = 0$ whereas in La$_2$NiO$_3$F$_2$ it is $k_x = 0$. 
This difference traces back to the in-phase vs antiphase anion-octahedra tiltings displayed by these two compounds. 

\begin{figure}[t!]
\includegraphics[width=.48\textwidth]{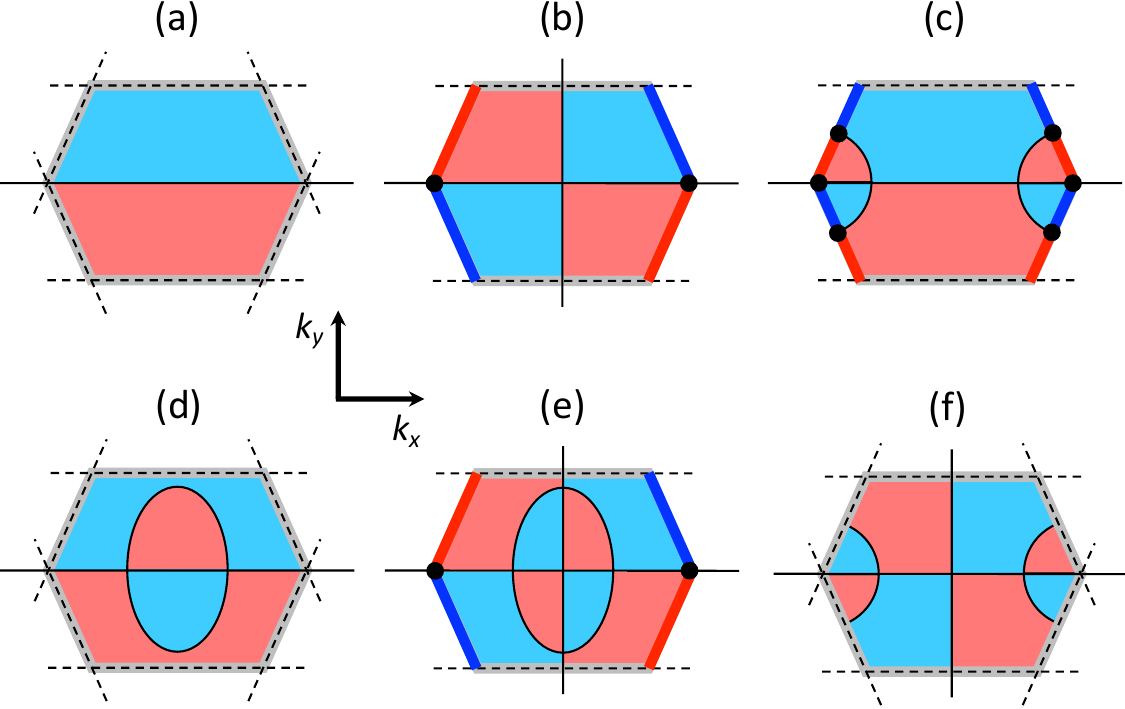}
    \caption{
    Top view of the first Brillouin zone illustrating different types of spin-momentum textures in Ruddledsen-Popper materials like La$_2$NiO$_4$ and La$_2$NiO$_3$F (red and blue indicate reversed spin splittings). 
    The vertical and horizontal solid lines indicate nodal planes imposed by symmetry ($k_x = 0$ and $k_y = 0$ respectively) while the curved lines represent accidental nodal surfaces.
    The nodal planes separate different sectors with broken spin degeneracy whose propagation to the second Brillouin zone is such that the zone boundary remains entirely nodal in (a) while it displays a partial breaking of the spin degeneracy in (b) [gray vs red and blue thick lines]. 
    This trivial vs non-trivial spin-momentum texture at the Brillouin-zone boundary survives in the presence of accidental nodes in (d) and (e).    
    In (b) and (e), the edges indicated by the black dots are nodal lines contained in only one nodal plane (differently from the nodal line at the zone center, which is contained in two planes). 
    In (c), the Brillouin-zone boundary displays additional ``isolated'' lines (black dots) that are nodal due to symmetry. 
    The pinning of accidental nodal surfaces at such ``isolated'' nodal lines yields a non-trivial spin-momentum texture at the boundary, which otherwise would remain as in (a) or (d). Note that this texture in (c) is different compared to (b). Conversely, the pining of accidental nodes to the zone boundary in (b) or (e) would restore the spin degeneracy as illustrated in (f). The spin-momentum textures of La$_2$NiO$_4$ and La$_2$NiO$_3$F correspond to (c) and (e) respectively, while the other panels illustrate additional possibilities that could be realized in other materials. 
    }
    \label{nodes}
\end{figure}

Interestingly, these two different situations would automatically translate into qualitatively different spin-momentum textures, not only within the Brillouin zone, but also at its boundary. 
This is illustrated in Fig.~\ref{nodes}. 
In the case of the two nodal planes at $k_y = 0$ and $k_z =0$ [Fig.~\ref{nodes}~(a)],  
the propagation of the spin splitting to the second Brillouin zone is such that the whole Brillouin-zone boundary should be nodal. 
In the case of the two nodal planes $k_y =0$ and $k_x = 0$ [Fig.~\ref{nodes}~(b)],
in contrast, the spin degeneracy needs to be broken at some parts of that boundary.

\begin{figure}[t!]
\includegraphics[width=.4\textwidth]{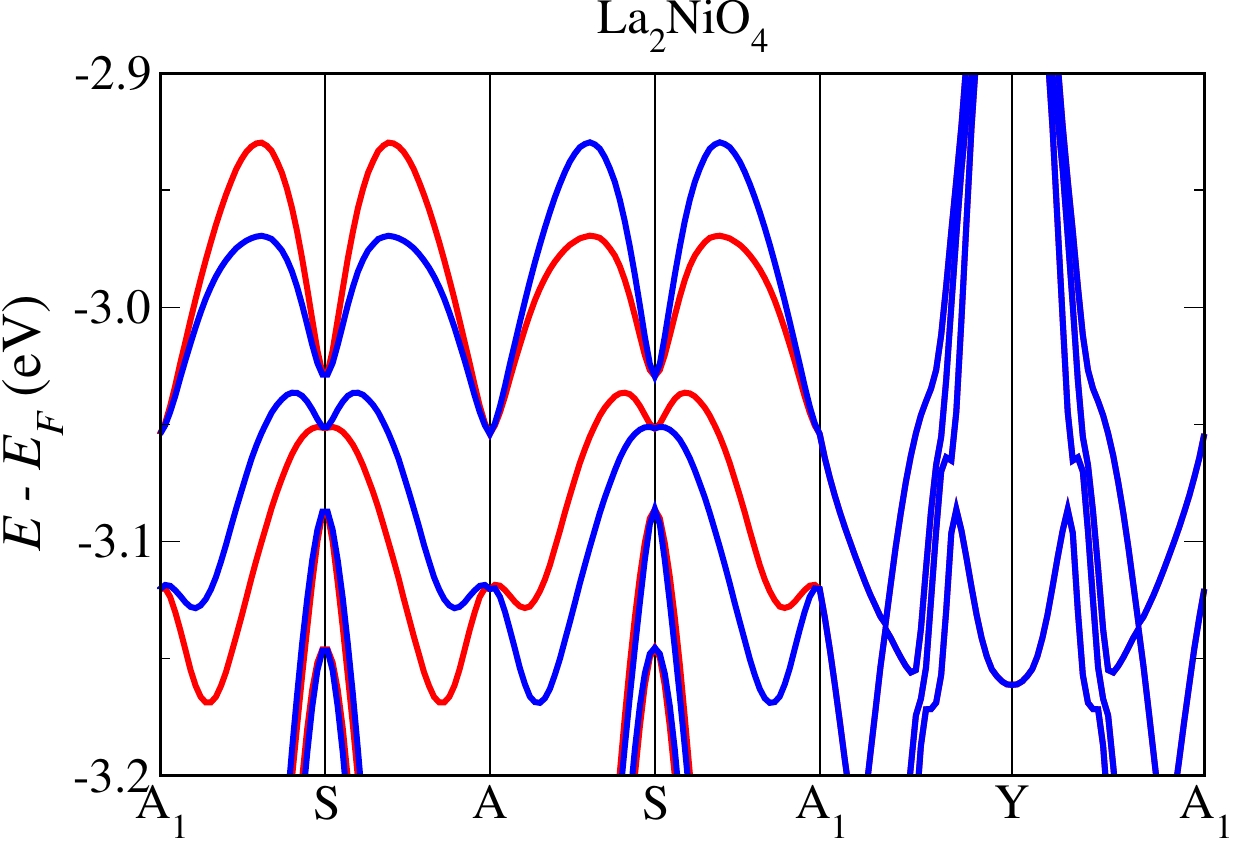}\\[1ex]
\includegraphics[width=.4\textwidth]{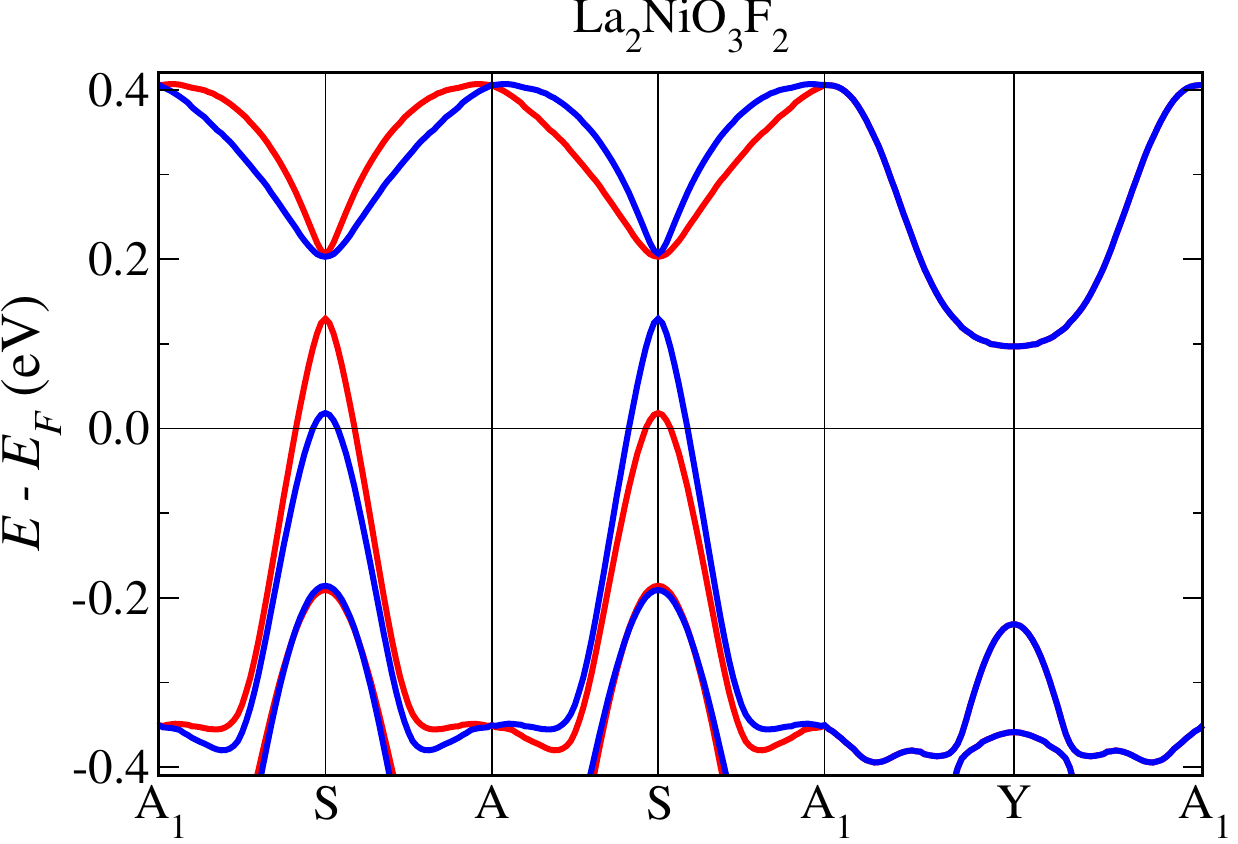}
\caption{
    Nonrelativistic, spin-projected band structure at the boundary of the Brillouin zone of the single-layer nickelates La$_2$NiO$_4$ and La$_2$NiO$_3$F$_2$
    in their collinear $G$-AFM magnetic state [see Fig. \ref{structure}(b)]. 
    These band plots illustrate that the splitting of the `up' and `down' (blue and red) bands associated with AM is protected in part of the Brillouin zone only. Note that the S point is nodal in La$_2$NiO$_4$ but not in La$_2$NiO$_3$F$_2$ (while the $A_1$-Y-$A_1$ path is nodal in both cases).  
    }
    \label{single-layer.bBZ}
\end{figure}

However, we find that the actual situation is a little bit more complex due to the presence of additional nodes. 
On one hand, we have the presence of accidental nodes as illustrated by the inset of Fig. \ref{single-layer}. 
These nodes appear in the form of curved surfaces that determine the eventual number of Brillouin-zone sectors with reversed spin splitting as sketched in Fig. \ref{nodes} (c)-(f). 
On the other hand, we note that there can be additional nodal lines at the boundary of the Brillouin zone which is eventually determined by overall magneto-structural symmetry of the system under consideration. 
This circumstance occurs in La$_2$NiO$_4$ along the R-S $k$-lines of its Brillouin zone.
Further, these lines turn out to pin the accidental nodes of this system in such a way that its Brillouin-zone boundary develops a non-trivial topology of spin-momentum texture. 
This is sketched in Fig. \ref{nodes} (c) and further illustrated in Fig. \ref{single-layer.bBZ}. 
This analysis reveals the emergence of different topologies of the spin-momentum texture at the boundary of the Brillouin zone, which can be used to refine the classification different AM materials (beyond $d$-wave or higher even-parity wave classes).   

\subsection{Bilayer %($n = 2$) 
AM nickelates}

As a representative of the bilayer ($n = 2$) Ruddlesden-Popper phases we consider the nickelate La$_3$Ni$_2$O$_7$. 
The crystal structure of this system corresponds to the $Amam$ space group with the Ni atoms at the $8g$ Wyckoff positions \cite{neumeier22}. 
Thus, the system displays rotations of the oxygen octahedra similar to the previous ones. Interestingly, the application of pressure suppresses these rotations and promotes high-temperature superconductivity with $T_c =80$~K \cite{wang23-bilayer}. 
When it comes to magnetism, the situation is less clear although experimental evidence of spin order at ambient pressure has been reported \cite{liu22}.

\begin{figure}[t!]
\includegraphics[width=.4\textwidth]{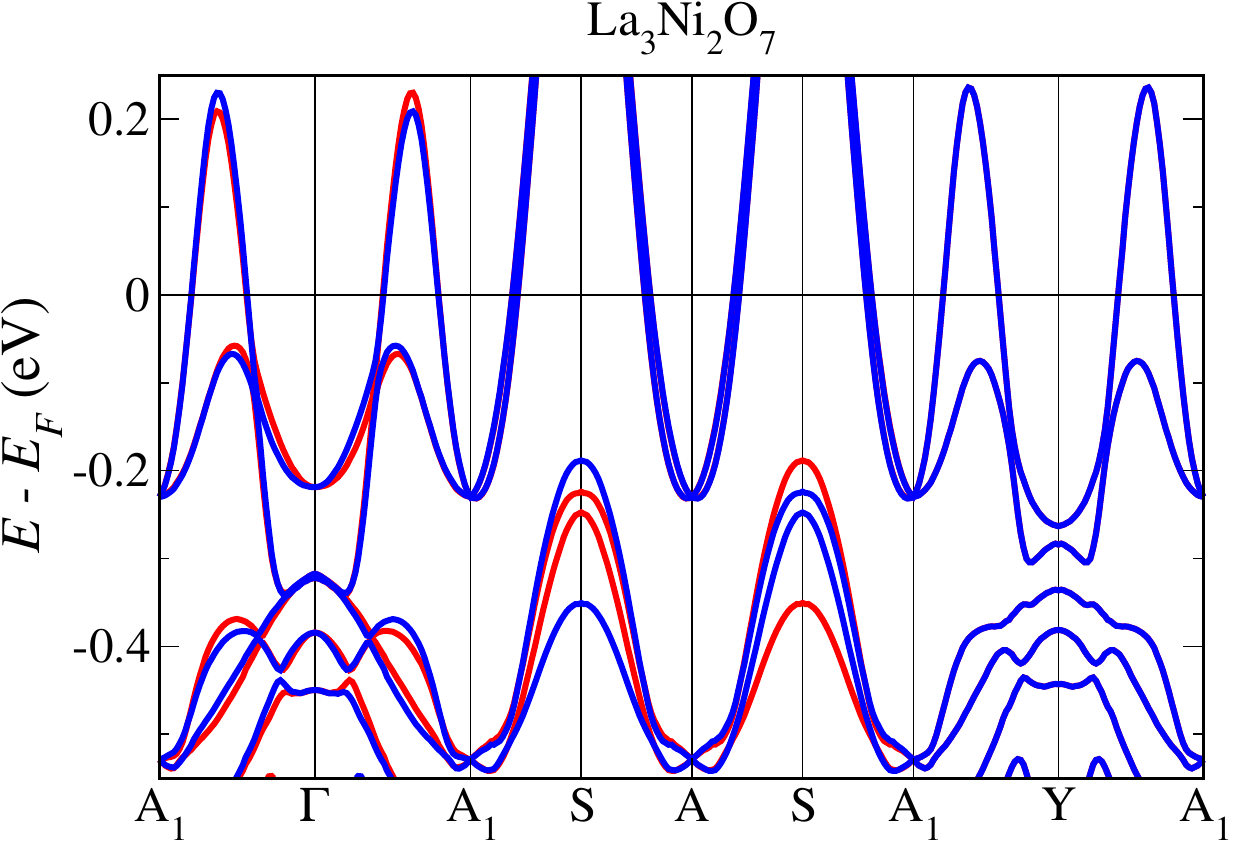}
    \caption{
    Nonrelativistic, spin-projected band structure of the bilayer nickelate La$_3$Ni$_2$O$_7$ with collinear $G$-AFM magnetic state along a path within the Brillouin zone and at its boundary. While the spin degeneracy is protected in part of the Brillouin-zone boundary, it is broken elsewhere as a manifestation of AM. 
    }
    \label{bilayer}
\end{figure}

Figure \ref{bilayer} shows the computed band structure of La$_2$Ni$_2$O$_7$ assuming the same type of AFM order as for the single-layer case, which can be the ground-state if correlations are weak enough (see also \cite{botana23}). The corresponding magnetic moment is $0.58$~$\mu_{\rm B}$ using the PBE functional, and the system remains metallic. 
The spin splitting of the bands is 28~meV near the Fermi level and reaches 48~meV further away. This splitting undergoes accidental nodes also, as illustrated with the Fermi surface shown in Table \ref{tab:summary}.%\\[-2.5em] 

\subsection{Perovskite %($n = \infty$) 
AM nickelates}

%\vspace{-.5em}
The $n = \infty$ end member of the Ruddlesden-Popper series corresponds to the perovskite structure. In this case, the distortion of the ideal cubic structure can generally be anticipated from the Goldschmidt tolerance factor. 
The rare-earth $R$NiO$_3$ nickelates belong to this class. These systems are well-known for their metal-insulator transition, which is accompanied by a structural distortion and either concomitant or subsequent AFM order. 
Specifically, they display a monoclinic $P2_1/n$ structure with tilted oxygen octahedra and two inequivalent Ni atoms at the 2$d$ and 2$c$ Wyckoff positions \cite{alonso01}. In this structure, $G$-type AFM order would also give rise to altermagnetism. These nickelates, however, display more complex magnetic configurations where the size of the magnetic unit cell increases compared with the crystallographic one. 
This circumstance implies conventional antiferromagnetism where spin degeneracy is preserved. 
Yet, the tendency of these systems towards $G$-AFM order may result into AM fluctuations (that is, deviations from the magnetic ground state that break spin degeneracy, even if this is restored in average).

Spin fluctuations are generally believed to be important for unconventional superconductivity. The latter has been observed in the reduced form of the above perovskite nickelates \cite{review-nickelates20}. These so-called infinite-layer nickelates also have a tendency to display structural distortions due to geometric effects \cite{cano22-prm,chen22-prb}. In the presence of these distortions, the theoretical magnetic ground state of these nickelates implies AM metallic behavior as illustrated for LaNiO$_2$ in Table \ref{tab:summary}. 

Besides the rare-earth $R$NiO$_3$ series and their reduced counterparts, there exist two other examples of $n = \infty$ nickelates whose synthesis, however, has been achieved by means of high pressure-high temperature processes only. The first one is BiNiO$_3$ \cite{kato02}. At ambient pressure, the crystal structure of this system corresponds to the triclinic $P\bar{1}$ space group with tilted oxygen octahedra. These tilts are such that there are four nonequivalent Ni sites within the unit cell. Theoretically, this enables imbalance of the otherwise AFM-ordered spins and therefore a ferromagnetic component of the total magnetization --- {\it i.e.} ferrimagnetism. 
However, the deviations of the individual moments from the average are in reality very small so that the magnetic configuration of this system can be approximated very well by assuming collinear $G$-AFM order with perfectly compensated spins \cite{attfield08}. Thus, BiNiO$_3$ provides an example of the Luttinger-compensated ferrimagnet discussed in \cite{mazin22-prx}, in which the ferromagnetic-like properties should be understood as due to altermagnetism (rather than due to its net magnetization, since it is virtually zero).
Interestingly, BiNiO$_3$ undergoes a $P\bar{1} \to Pbnm$ structural transition under pressure \cite{attfield07}. This transition changes the conduction type from insulating to metallic so that the AM behavior of this system may be supplemented with metallicity. In that case, the computed spin splitting reaches 167 meV.

The second example is PbNiO$_3$ \cite{saitoh11}. In this case, the crystal structure corresponds to the orthorhombic $Pnma$ space group with the Ni atoms in the 4$a$ Wyckoff positions. $G$-AFM order has been predicted for this system \cite{franchini12-prb, picozzi14}, which therefore should be AM as well. In fact, in our calculations, the computed spin splitting reaches 334 meV and its symmetry is illustrated in Table \ref{tab:summary}.

In addition, the crystal structure of PbNiO$_3$ can be transformed into a rhombohedrally distorted $R3c$ perovskite structure by heat treatment. This structure lacks inversion symmetry. Further, the primitive $R3c$ unit cell contains two equivalent Ni-atom positions and this cell corresponds to the magnetic unit cell for the corresponding $G$-AFM order. Consequently, PbNiO$_3$ is AM in this $R3c$ structure also.%\\[-2.25em]

\section{Additional materials}

%\vspace{-.5em}

In the following, we further argue that Ruddledsen-Popper and perovskite phases are particularly favorable setups for AM behavior by pointing out specific examples beyond the Ni-based materials discussed above. 
We note that two analogs of La$_2$NiO$_4$ have previously been proposed as AM materials, the single-layer cuprate La$_2$CuO$_4$ \cite{jungwirth22-prx} and the single-layer manganite Ca$_2$MnO$_4$ \cite{zunger22}. 
In the case of Mn-based AFM materials of this type, the tilting of the oxygen octahedra enabling AM features can also be obtained in (Sr,La)$_2$MnO$_4$ \cite{shimono06}. 
Further, the crystal structure of the single-layer La$_2$CoO$_4$ compounds also displays similar oxygen-octahedra tiltings, and this system has two successive AFM transitions at 275~K and 135~K that make this material the Co-based analog of La$_2$NiO$_4$ and La$_2$CuO$_4$, respectively \cite{shirane89}.
The Cr-based systems Ca$R$CrO$_4$ ($R = $Pr, Nd, Sm and Eu) are also remarkable analogs of the single-layer AM prototype La$_2$NiO$_4$ \cite{martinez1,martinez2,martinez3}. 
Specifically, these systems have the same $Bmab$ crystal structure with tilted oxygen octahedra and the same type of $G$-AFM order emerging at $\sim$200~K. 
These single-layer Ruddledsen-Popper phases thus expand the material base of AM systems.  

When it comes to the $n=\infty$ end members of the Ruddledsen-Popper series, we note that the most widely studied multiferroic material, BiFeO$_3$, also hosts AM properties under the appropriate circumstances. 
The crystal structure of this system corresponds to the rhombohedrally distorted $R3c$ perovskite with the Fe atoms at the 6$a$ Wyckoff positions. AFM order of the $G$ type with the spins along the [111] pseudocubic direction was initially reported for this system. Thus, similarly to PbNiO$_3$ in its $R3c$ phase, such a $G$-AFM order would imply AM behavior in BiFeO$_3$ also. 
In that case, the spin splitting would reach 227~meV and its symmetry is illustrated in Table \ref{tab:summary}. 
These features, however, are washed out by the spiral modulation of the $G$-AFM that results from relativistic effects \cite{steichele82}. 
At the same time, the period of the spiral is very long ($\sim 62$~nm). Consequently, the spiral modulation can effectively be truncated in thin films \cite{sando20} and thereby the underlying altermagnetism of the multiferroic BiFeO$_3$ can be restored.%\\[-2.25em]

\section{Conclusions}

%\vspace{-.75em}
We have shown that the Ruddlesden-Popper phases, including their perovskite end members, extensively provide altermagnetic materials. 
The reason is that collinear antiferromagnetic order very frequently occurs in these systems in the presence of structural distortions that prevent the transformation of opposite-spin sublattices into each other via the symmetry operations of inversion or translation. 
Besides the realization of altermagnetism itself, our analysis has revealed the generic presence of accidental nodes and topologically non-trivial spin-momentum textures at the boundary of the Brillouin zone. These features can in principle be used to refine the classification of different types of altermagnetic order (beyond $d$-wave, $g$-wave, etc.).
This circumstance has been illustrated for prototypical Ni-based materials as well as for systems with other transition metals (Cu, Co, Fe, Mn, and Cr), where altermagnetism may supplement other properties such as superconductivity and multiferroicity. 

Among the materials that we have considered, BiNiO$_3$ and the prototypical multiferroic BiFeO$_3$ appear as particularly interesting systems in relation to the classification of different types of magnetic order based on measurable properties. 
BiNiO$_3$ formally displays a ferrimagnetic order.
However, following \cite{mazin22-prx}, we have argued that it should be considered as a Luttinger-compensated ferrimagnet with ferromagnetic-like properties emerging from the purely antiferromagnetic component of its magnetic order via altermagnetism. 
Further, by analogy from this example, we anticipate the existence of weak ferromagnets whose ferromagnetic-like properties do not result from the ferromagnetic component of its total magnetic order in first place, but from the antiferromagnetic one instead via altermagnetism. This interplay has recently been discussed in relation to RF$_4$ for example \cite{Milivojević_2024} (see also \cite{lee24}).
According to this observation, altermagnetism then represents a new fundamental ingredient that challenges and at the same time enriches the formal classes of ferro-, ferri-, and antiferro-magnets as well as their boundaries. BiFeO$_3$, in its turn, may be regarded as a failed altermagnet due to the spiral modulation of its antiferromagnetic order. 
This modulation, however, may be neutralized under the appropriate conditions, which should enable the emergence and control of altermagnetic behavior and hence ferromagnetic-like properties in this multiferroic system. 
These examples additionally show that altermagnetism can also be investigated and put to use in systems beyond the initially proposed ones (i.e. beyond collinear antiferromagnets with perfectly compensated magnetization).
Thus, we expect that our findings will motivate further work, both theoretical and experimental, on the new altermagnetic perspective to magnetism. 

%\printbibliography
\bibliography{bib.bib}

\begin{thebibliography}{59}%
\makeatletter
\providecommand \@ifxundefined [1]{%
 \@ifx{#1\undefined}
}%
\providecommand \@ifnum [1]{%
 \ifnum #1\expandafter \@firstoftwo
 \else \expandafter \@secondoftwo
 \fi
}%
\providecommand \@ifx [1]{%
 \ifx #1\expandafter \@firstoftwo
 \else \expandafter \@secondoftwo
 \fi
}%
\providecommand \natexlab [1]{#1}%
\providecommand \enquote  [1]{``#1''}%
\providecommand \bibnamefont  [1]{#1}%
\providecommand \bibfnamefont [1]{#1}%
\providecommand \citenamefont [1]{#1}%
\providecommand \href@noop [0]{\@secondoftwo}%
\providecommand \href [0]{\begingroup \@sanitize@url \@href}%
\providecommand \@href[1]{\@@startlink{#1}\@@href}%
\providecommand \@@href[1]{\endgroup#1\@@endlink}%
\providecommand \@sanitize@url [0]{\catcode `\\12\catcode `\$12\catcode
  `\&12\catcode `\#12\catcode `\^12\catcode `\_12\catcode `\%12\relax}%
\providecommand \@@startlink[1]{}%
\providecommand \@@endlink[0]{}%
\providecommand \url  [0]{\begingroup\@sanitize@url \@url }%
\providecommand \@url [1]{\endgroup\@href {#1}{\urlprefix }}%
\providecommand \urlprefix  [0]{URL }%
\providecommand \Eprint [0]{\href }%
\providecommand \doibase [0]{http://dx.doi.org/}%
\providecommand \selectlanguage [0]{\@gobble}%
\providecommand \bibinfo  [0]{\@secondoftwo}%
\providecommand \bibfield  [0]{\@secondoftwo}%
\providecommand \translation [1]{[#1]}%
\providecommand \BibitemOpen [0]{}%
\providecommand \bibitemStop [0]{}%
\providecommand \bibitemNoStop [0]{.\EOS\space}%
\providecommand \EOS [0]{\spacefactor3000\relax}%
\providecommand \BibitemShut  [1]{\csname bibitem#1\endcsname}%
\let\auto@bib@innerbib\@empty
%</preamble>
\bibitem [{\citenamefont {Hayami}\ \emph {et~al.}(2019)\citenamefont {Hayami},
  \citenamefont {Yanagi},\ and\ \citenamefont {Kusunose}}]{kusunose19}%
  \BibitemOpen
  \bibfield  {author} {\bibinfo {author} {\bibfnamefont {S.}~\bibnamefont
  {Hayami}}, \bibinfo {author} {\bibfnamefont {Y.}~\bibnamefont {Yanagi}}, \
  and\ \bibinfo {author} {\bibfnamefont {H.}~\bibnamefont {Kusunose}},\ }\href
  {\doibase 10.7566/JPSJ.88.123702} {\bibfield  {journal} {\bibinfo  {journal}
  {J. Phys. Soc. Jpn.}\ }\textbf {\bibinfo {volume} {88}},\ \bibinfo {pages}
  {123702} (\bibinfo {year} {2019})}\BibitemShut {NoStop}%
\bibitem [{\citenamefont {Ahn}\ \emph {et~al.}(2019)\citenamefont {Ahn},
  \citenamefont {Hariki}, \citenamefont {Lee},\ and\ \citenamefont
  {Kune\ifmmode~\check{s}\else \v{s}\fi{}}}]{ahn19}%
  \BibitemOpen
  \bibfield  {author} {\bibinfo {author} {\bibfnamefont {K.-H.}\ \bibnamefont
  {Ahn}}, \bibinfo {author} {\bibfnamefont {A.}~\bibnamefont {Hariki}},
  \bibinfo {author} {\bibfnamefont {K.-W.}\ \bibnamefont {Lee}}, \ and\
  \bibinfo {author} {\bibfnamefont {J.}~\bibnamefont
  {Kune\ifmmode~\check{s}\else \v{s}\fi{}}},\ }\href {\doibase
  10.1103/PhysRevB.99.184432} {\bibfield  {journal} {\bibinfo  {journal} {Phys.
  Rev. B}\ }\textbf {\bibinfo {volume} {99}},\ \bibinfo {pages} {184432}
  (\bibinfo {year} {2019})}\BibitemShut {NoStop}%
\bibitem [{\citenamefont {Naka}\ \emph {et~al.}(2019)\citenamefont {Naka},
  \citenamefont {Hayami}, \citenamefont {Kusunose}, \citenamefont {Yanagi},
  \citenamefont {Motome},\ and\ \citenamefont {Seo}}]{Naka2019}%
  \BibitemOpen
  \bibfield  {author} {\bibinfo {author} {\bibfnamefont {M.}~\bibnamefont
  {Naka}}, \bibinfo {author} {\bibfnamefont {S.}~\bibnamefont {Hayami}},
  \bibinfo {author} {\bibfnamefont {H.}~\bibnamefont {Kusunose}}, \bibinfo
  {author} {\bibfnamefont {Y.}~\bibnamefont {Yanagi}}, \bibinfo {author}
  {\bibfnamefont {Y.}~\bibnamefont {Motome}}, \ and\ \bibinfo {author}
  {\bibfnamefont {H.}~\bibnamefont {Seo}},\ }\href {\doibase
  10.1038/s41467-019-12229-y} {\bibfield  {journal} {\bibinfo  {journal}
  {Nature Comm.}\ }\textbf {\bibinfo {volume} {10}},\ \bibinfo {pages} {4305}
  (\bibinfo {year} {2019})}\BibitemShut {NoStop}%
\bibitem [{\citenamefont {Šmejkal}\ \emph {et~al.}(2020)\citenamefont
  {Šmejkal}, \citenamefont {González-Hernández}, \citenamefont {Jungwirth},\
  and\ \citenamefont {Sinova}}]{smejkal20}%
  \BibitemOpen
  \bibfield  {author} {\bibinfo {author} {\bibfnamefont {L.}~\bibnamefont
  {Šmejkal}}, \bibinfo {author} {\bibfnamefont {R.}~\bibnamefont
  {González-Hernández}}, \bibinfo {author} {\bibfnamefont {T.}~\bibnamefont
  {Jungwirth}}, \ and\ \bibinfo {author} {\bibfnamefont {J.}~\bibnamefont
  {Sinova}},\ }\href {\doibase 10.1126/sciadv.aaz8809} {\bibfield  {journal}
  {\bibinfo  {journal} {Science Advances}\ }\textbf {\bibinfo {volume} {6}},\
  \bibinfo {pages} {eaaz8809} (\bibinfo {year} {2020})}\BibitemShut {NoStop}%
\bibitem [{\citenamefont {Samanta}\ \emph {et~al.}(2020)\citenamefont
  {Samanta}, \citenamefont {Ležaić}, \citenamefont {Merte}, \citenamefont
  {Freimuth}, \citenamefont {Blügel},\ and\ \citenamefont
  {Mokrousov}}]{mokrousov20}%
  \BibitemOpen
  \bibfield  {author} {\bibinfo {author} {\bibfnamefont {K.}~\bibnamefont
  {Samanta}}, \bibinfo {author} {\bibfnamefont {M.}~\bibnamefont {Ležaić}},
  \bibinfo {author} {\bibfnamefont {M.}~\bibnamefont {Merte}}, \bibinfo
  {author} {\bibfnamefont {F.}~\bibnamefont {Freimuth}}, \bibinfo {author}
  {\bibfnamefont {S.}~\bibnamefont {Blügel}}, \ and\ \bibinfo {author}
  {\bibfnamefont {Y.}~\bibnamefont {Mokrousov}},\ }\href {\doibase
  10.1063/5.0005017} {\bibfield  {journal} {\bibinfo  {journal} {J. Appl.
  Phys.}\ }\textbf {\bibinfo {volume} {127}},\ \bibinfo {pages} {213904}
  (\bibinfo {year} {2020})}\BibitemShut {NoStop}%
\bibitem [{\citenamefont {Yuan}\ \emph {et~al.}(2020)\citenamefont {Yuan},
  \citenamefont {Wang}, \citenamefont {Luo}, \citenamefont {Rashba},\ and\
  \citenamefont {Zunger}}]{zunger20-prb}%
  \BibitemOpen
  \bibfield  {author} {\bibinfo {author} {\bibfnamefont {L.-D.}\ \bibnamefont
  {Yuan}}, \bibinfo {author} {\bibfnamefont {Z.}~\bibnamefont {Wang}}, \bibinfo
  {author} {\bibfnamefont {J.-W.}\ \bibnamefont {Luo}}, \bibinfo {author}
  {\bibfnamefont {E.~I.}\ \bibnamefont {Rashba}}, \ and\ \bibinfo {author}
  {\bibfnamefont {A.}~\bibnamefont {Zunger}},\ }\href {\doibase
  10.1103/PhysRevB.102.014422} {\bibfield  {journal} {\bibinfo  {journal}
  {Phys. Rev. B}\ }\textbf {\bibinfo {volume} {102}},\ \bibinfo {pages}
  {014422} (\bibinfo {year} {2020})}\BibitemShut {NoStop}%
\bibitem [{\citenamefont {Feng}\ \emph {et~al.}(2022)\citenamefont {Feng},
  \citenamefont {Zhou}, \citenamefont {{\v{S}}mejkal}, \citenamefont {Wu},
  \citenamefont {Zhu}, \citenamefont {Guo}, \citenamefont
  {Gonz{\'a}lez-Hern{\'a}ndez}, \citenamefont {Wang}, \citenamefont {Yan},
  \citenamefont {Qin}, \citenamefont {Zhang}, \citenamefont {Wu}, \citenamefont
  {Chen}, \citenamefont {Meng}, \citenamefont {Liu}, \citenamefont {Xia},
  \citenamefont {Sinova}, \citenamefont {Jungwirth},\ and\ \citenamefont
  {Liu}}]{Feng2022}%
  \BibitemOpen
  \bibfield  {author} {\bibinfo {author} {\bibfnamefont {Z.}~\bibnamefont
  {Feng}}, \bibinfo {author} {\bibfnamefont {X.}~\bibnamefont {Zhou}}, \bibinfo
  {author} {\bibfnamefont {L.}~\bibnamefont {{\v{S}}mejkal}}, \bibinfo {author}
  {\bibfnamefont {L.}~\bibnamefont {Wu}}, \bibinfo {author} {\bibfnamefont
  {Z.}~\bibnamefont {Zhu}}, \bibinfo {author} {\bibfnamefont {H.}~\bibnamefont
  {Guo}}, \bibinfo {author} {\bibfnamefont {R.}~\bibnamefont
  {Gonz{\'a}lez-Hern{\'a}ndez}}, \bibinfo {author} {\bibfnamefont
  {X.}~\bibnamefont {Wang}}, \bibinfo {author} {\bibfnamefont {H.}~\bibnamefont
  {Yan}}, \bibinfo {author} {\bibfnamefont {P.}~\bibnamefont {Qin}}, \bibinfo
  {author} {\bibfnamefont {X.}~\bibnamefont {Zhang}}, \bibinfo {author}
  {\bibfnamefont {H.}~\bibnamefont {Wu}}, \bibinfo {author} {\bibfnamefont
  {H.}~\bibnamefont {Chen}}, \bibinfo {author} {\bibfnamefont {Z.}~\bibnamefont
  {Meng}}, \bibinfo {author} {\bibfnamefont {L.}~\bibnamefont {Liu}}, \bibinfo
  {author} {\bibfnamefont {Z.}~\bibnamefont {Xia}}, \bibinfo {author}
  {\bibfnamefont {J.}~\bibnamefont {Sinova}}, \bibinfo {author} {\bibfnamefont
  {T.}~\bibnamefont {Jungwirth}}, \ and\ \bibinfo {author} {\bibfnamefont
  {Z.}~\bibnamefont {Liu}},\ }\href {\doibase 10.1038/s41928-022-00866-z}
  {\bibfield  {journal} {\bibinfo  {journal} {Nature Electronics}\ }\textbf
  {\bibinfo {volume} {5}},\ \bibinfo {pages} {735} (\bibinfo {year}
  {2022})}\BibitemShut {NoStop}%
\bibitem [{\citenamefont {Cuono}\ \emph {et~al.}(2023)\citenamefont {Cuono},
  \citenamefont {Sattigeri}, \citenamefont {Skolimowski},\ and\ \citenamefont
  {Autieri}}]{CUONO2023171163}%
  \BibitemOpen
  \bibfield  {author} {\bibinfo {author} {\bibfnamefont {G.}~\bibnamefont
  {Cuono}}, \bibinfo {author} {\bibfnamefont {R.~M.}\ \bibnamefont
  {Sattigeri}}, \bibinfo {author} {\bibfnamefont {J.}~\bibnamefont
  {Skolimowski}}, \ and\ \bibinfo {author} {\bibfnamefont {C.}~\bibnamefont
  {Autieri}},\ }\href {\doibase https://doi.org/10.1016/j.jmmm.2023.171163}
  {\bibfield  {journal} {\bibinfo  {journal} {Journal of Magnetism and Magnetic
  Materials}\ }\textbf {\bibinfo {volume} {586}},\ \bibinfo {pages} {171163}
  (\bibinfo {year} {2023})}\BibitemShut {NoStop}%
\bibitem [{\citenamefont {Yuan}\ and\ \citenamefont {Zunger}(2023)}]{zunger23}%
  \BibitemOpen
  \bibfield  {author} {\bibinfo {author} {\bibfnamefont {L.-D.}\ \bibnamefont
  {Yuan}}\ and\ \bibinfo {author} {\bibfnamefont {A.}~\bibnamefont {Zunger}},\
  }\href {\doibase https://doi.org/10.1002/adma.202211966} {\bibfield
  {journal} {\bibinfo  {journal} {Advanced Materials}\ }\textbf {\bibinfo
  {volume} {35}},\ \bibinfo {pages} {2211966} (\bibinfo {year}
  {2023})}\BibitemShut {NoStop}%
\bibitem [{\citenamefont {Guo}\ \emph {et~al.}(2023{\natexlab{a}})\citenamefont
  {Guo}, \citenamefont {Liu}, \citenamefont {Janson}, \citenamefont {Fulga},
  \citenamefont {{van den Brink}},\ and\ \citenamefont
  {Facio}}]{GUO2023100991}%
  \BibitemOpen
  \bibfield  {author} {\bibinfo {author} {\bibfnamefont {Y.}~\bibnamefont
  {Guo}}, \bibinfo {author} {\bibfnamefont {H.}~\bibnamefont {Liu}}, \bibinfo
  {author} {\bibfnamefont {O.}~\bibnamefont {Janson}}, \bibinfo {author}
  {\bibfnamefont {I.~C.}\ \bibnamefont {Fulga}}, \bibinfo {author}
  {\bibfnamefont {J.}~\bibnamefont {{van den Brink}}}, \ and\ \bibinfo {author}
  {\bibfnamefont {J.~I.}\ \bibnamefont {Facio}},\ }\href {\doibase
  https://doi.org/10.1016/j.mtphys.2023.100991} {\bibfield  {journal} {\bibinfo
   {journal} {Materials Today Physics}\ }\textbf {\bibinfo {volume} {32}},\
  \bibinfo {pages} {100991} (\bibinfo {year} {2023}{\natexlab{a}})}\BibitemShut
  {NoStop}%
\bibitem [{\citenamefont {Chen}\ \emph {et~al.}(2023)\citenamefont {Chen},
  \citenamefont {Wang}, \citenamefont {Li},\ and\ \citenamefont
  {Sanyal}}]{ang23a}%
  \BibitemOpen
  \bibfield  {author} {\bibinfo {author} {\bibfnamefont {X.}~\bibnamefont
  {Chen}}, \bibinfo {author} {\bibfnamefont {D.}~\bibnamefont {Wang}}, \bibinfo
  {author} {\bibfnamefont {L.}~\bibnamefont {Li}}, \ and\ \bibinfo {author}
  {\bibfnamefont {B.}~\bibnamefont {Sanyal}},\ }\href {\doibase
  10.1063/5.0147450} {\bibfield  {journal} {\bibinfo  {journal} {Applied
  Physics Letters}\ }\textbf {\bibinfo {volume} {123}},\ \bibinfo {pages}
  {022402} (\bibinfo {year} {2023})}\BibitemShut {NoStop}%
\bibitem [{\citenamefont {Guo}\ and\ \citenamefont {Ang}(2023)}]{ang23b}%
  \BibitemOpen
  \bibfield  {author} {\bibinfo {author} {\bibfnamefont {S.-D.}\ \bibnamefont
  {Guo}}\ and\ \bibinfo {author} {\bibfnamefont {Y.~S.}\ \bibnamefont {Ang}},\
  }\href {\doibase 10.1103/PhysRevB.108.L180403} {\bibfield  {journal}
  {\bibinfo  {journal} {Phys. Rev. B}\ }\textbf {\bibinfo {volume} {108}},\
  \bibinfo {pages} {L180403} (\bibinfo {year} {2023})}\BibitemShut {NoStop}%
\bibitem [{\citenamefont {Guo}\ \emph {et~al.}(2023{\natexlab{b}})\citenamefont
  {Guo}, \citenamefont {Guo}, \citenamefont {Cheng}, \citenamefont {Wang},\
  and\ \citenamefont {Ang}}]{ang23c}%
  \BibitemOpen
  \bibfield  {author} {\bibinfo {author} {\bibfnamefont {S.-D.}\ \bibnamefont
  {Guo}}, \bibinfo {author} {\bibfnamefont {X.-S.}\ \bibnamefont {Guo}},
  \bibinfo {author} {\bibfnamefont {K.}~\bibnamefont {Cheng}}, \bibinfo
  {author} {\bibfnamefont {K.}~\bibnamefont {Wang}}, \ and\ \bibinfo {author}
  {\bibfnamefont {Y.~S.}\ \bibnamefont {Ang}},\ }\href {\doibase
  10.1063/5.0161431} {\bibfield  {journal} {\bibinfo  {journal} {Appl. Phys.
  Lett.}\ }\textbf {\bibinfo {volume} {123}},\ \bibinfo {pages} {082401}
  (\bibinfo {year} {2023}{\natexlab{b}})}\BibitemShut {NoStop}%
\bibitem [{\citenamefont {{Gao}}\ \emph {et~al.}()\citenamefont {{Gao}},
  \citenamefont {{Qu}}, \citenamefont {{Zeng}}, \citenamefont {{Liu}},
  \citenamefont {{Wen}}, \citenamefont {{Sun}}, \citenamefont {{Guo}},\ and\
  \citenamefont {{Lu}}}]{gao23a}%
  \BibitemOpen
  \bibfield  {author} {\bibinfo {author} {\bibfnamefont {Z.-F.}\ \bibnamefont
  {{Gao}}}, \bibinfo {author} {\bibfnamefont {S.}~\bibnamefont {{Qu}}},
  \bibinfo {author} {\bibfnamefont {B.}~\bibnamefont {{Zeng}}}, \bibinfo
  {author} {\bibfnamefont {Y.}~\bibnamefont {{Liu}}}, \bibinfo {author}
  {\bibfnamefont {J.-R.}\ \bibnamefont {{Wen}}}, \bibinfo {author}
  {\bibfnamefont {H.}~\bibnamefont {{Sun}}}, \bibinfo {author} {\bibfnamefont
  {P.-J.}\ \bibnamefont {{Guo}}}, \ and\ \bibinfo {author} {\bibfnamefont
  {Z.-Y.}\ \bibnamefont {{Lu}}},\ }\href@noop {} {\ }\Eprint
  {http://arxiv.org/abs/2311.04418} {arXiv:2311.04418} \BibitemShut {NoStop}%
\bibitem [{\citenamefont {{Guo}}\ \emph {et~al.}()\citenamefont {{Guo}},
  \citenamefont {{Gu}}, \citenamefont {{Gao}},\ and\ \citenamefont
  {{Lu}}}]{gao23b}%
  \BibitemOpen
  \bibfield  {author} {\bibinfo {author} {\bibfnamefont {P.-J.}\ \bibnamefont
  {{Guo}}}, \bibinfo {author} {\bibfnamefont {Y.}~\bibnamefont {{Gu}}},
  \bibinfo {author} {\bibfnamefont {Z.-F.}\ \bibnamefont {{Gao}}}, \ and\
  \bibinfo {author} {\bibfnamefont {Z.-Y.}\ \bibnamefont {{Lu}}},\ }\href@noop
  {} {\ }\Eprint {http://arxiv.org/abs/2312.13911} {arXiv:2312.13911}
  \BibitemShut {NoStop}%
\bibitem [{\citenamefont {{Qu}}\ \emph {et~al.}()\citenamefont {{Qu}},
  \citenamefont {{Gao}}, \citenamefont {{Sun}}, \citenamefont {{Liu}},
  \citenamefont {{Guo}},\ and\ \citenamefont {{Lu}}}]{gao24}%
  \BibitemOpen
  \bibfield  {author} {\bibinfo {author} {\bibfnamefont {S.}~\bibnamefont
  {{Qu}}}, \bibinfo {author} {\bibfnamefont {Z.-F.}\ \bibnamefont {{Gao}}},
  \bibinfo {author} {\bibfnamefont {H.}~\bibnamefont {{Sun}}}, \bibinfo
  {author} {\bibfnamefont {K.}~\bibnamefont {{Liu}}}, \bibinfo {author}
  {\bibfnamefont {P.-J.}\ \bibnamefont {{Guo}}}, \ and\ \bibinfo {author}
  {\bibfnamefont {Z.-Y.}\ \bibnamefont {{Lu}}},\ }\href@noop {} {\ }\Eprint
  {http://arxiv.org/abs/2401.11065} {arXiv:2401.11065} \BibitemShut {NoStop}%
\bibitem [{\citenamefont {Fernandes}\ \emph {et~al.}(2024)\citenamefont
  {Fernandes}, \citenamefont {de~Carvalho}, \citenamefont {Birol},\ and\
  \citenamefont {Pereira}}]{fernandes24-prb}%
  \BibitemOpen
  \bibfield  {author} {\bibinfo {author} {\bibfnamefont {R.~M.}\ \bibnamefont
  {Fernandes}}, \bibinfo {author} {\bibfnamefont {V.~S.}\ \bibnamefont
  {de~Carvalho}}, \bibinfo {author} {\bibfnamefont {T.}~\bibnamefont {Birol}},
  \ and\ \bibinfo {author} {\bibfnamefont {R.~G.}\ \bibnamefont {Pereira}},\
  }\href {\doibase 10.1103/PhysRevB.109.024404} {\bibfield  {journal} {\bibinfo
   {journal} {Phys. Rev. B}\ }\textbf {\bibinfo {volume} {109}},\ \bibinfo
  {pages} {024404} (\bibinfo {year} {2024})}\BibitemShut {NoStop}%
\bibitem [{\citenamefont {Osumi}\ \emph {et~al.}(2024)\citenamefont {Osumi},
  \citenamefont {Souma}, \citenamefont {Aoyama}, \citenamefont {Yamauchi},
  \citenamefont {Honma}, \citenamefont {Nakayama}, \citenamefont {Takahashi},
  \citenamefont {Ohgushi},\ and\ \citenamefont {Sato}}]{sato24-prb}%
  \BibitemOpen
  \bibfield  {author} {\bibinfo {author} {\bibfnamefont {T.}~\bibnamefont
  {Osumi}}, \bibinfo {author} {\bibfnamefont {S.}~\bibnamefont {Souma}},
  \bibinfo {author} {\bibfnamefont {T.}~\bibnamefont {Aoyama}}, \bibinfo
  {author} {\bibfnamefont {K.}~\bibnamefont {Yamauchi}}, \bibinfo {author}
  {\bibfnamefont {A.}~\bibnamefont {Honma}}, \bibinfo {author} {\bibfnamefont
  {K.}~\bibnamefont {Nakayama}}, \bibinfo {author} {\bibfnamefont
  {T.}~\bibnamefont {Takahashi}}, \bibinfo {author} {\bibfnamefont
  {K.}~\bibnamefont {Ohgushi}}, \ and\ \bibinfo {author} {\bibfnamefont
  {T.}~\bibnamefont {Sato}},\ }\href {\doibase 10.1103/PhysRevB.109.115102}
  {\bibfield  {journal} {\bibinfo  {journal} {Phys. Rev. B}\ }\textbf {\bibinfo
  {volume} {109}},\ \bibinfo {pages} {115102} (\bibinfo {year}
  {2024})}\BibitemShut {NoStop}%
\bibitem [{\citenamefont {\ifmmode~\check{S}\else \v{S}\fi{}mejkal}\ \emph
  {et~al.}(2022)\citenamefont {\ifmmode~\check{S}\else \v{S}\fi{}mejkal},
  \citenamefont {Sinova},\ and\ \citenamefont {Jungwirth}}]{jungwirth22-prx}%
  \BibitemOpen
  \bibfield  {author} {\bibinfo {author} {\bibfnamefont {L.}~\bibnamefont
  {\ifmmode~\check{S}\else \v{S}\fi{}mejkal}}, \bibinfo {author} {\bibfnamefont
  {J.}~\bibnamefont {Sinova}}, \ and\ \bibinfo {author} {\bibfnamefont
  {T.}~\bibnamefont {Jungwirth}},\ }\href {\doibase 10.1103/PhysRevX.12.031042}
  {\bibfield  {journal} {\bibinfo  {journal} {Phys. Rev. X}\ }\textbf {\bibinfo
  {volume} {12}},\ \bibinfo {pages} {031042} (\bibinfo {year}
  {2022})}\BibitemShut {NoStop}%
\bibitem [{\citenamefont {Mazin}(2022)}]{mazin22-prx}%
  \BibitemOpen
  \bibfield  {author} {\bibinfo {author} {\bibfnamefont {I.}~\bibnamefont
  {Mazin}},\ }\href {\doibase 10.1103/PhysRevX.12.040002} {\bibfield  {journal}
  {\bibinfo  {journal} {Phys. Rev. X}\ }\textbf {\bibinfo {volume} {12}},\
  \bibinfo {pages} {040002} (\bibinfo {year} {2022})}\BibitemShut {NoStop}%
\bibitem [{\citenamefont {Bonbien}\ \emph {et~al.}(2021)\citenamefont
  {Bonbien}, \citenamefont {Zhuo}, \citenamefont {Salimath}, \citenamefont
  {Ly}, \citenamefont {Abbout},\ and\ \citenamefont
  {Manchon}}]{manchon22-jpd:ap}%
  \BibitemOpen
  \bibfield  {author} {\bibinfo {author} {\bibfnamefont {V.}~\bibnamefont
  {Bonbien}}, \bibinfo {author} {\bibfnamefont {F.}~\bibnamefont {Zhuo}},
  \bibinfo {author} {\bibfnamefont {A.}~\bibnamefont {Salimath}}, \bibinfo
  {author} {\bibfnamefont {O.}~\bibnamefont {Ly}}, \bibinfo {author}
  {\bibfnamefont {A.}~\bibnamefont {Abbout}}, \ and\ \bibinfo {author}
  {\bibfnamefont {A.}~\bibnamefont {Manchon}},\ }\href {\doibase
  10.1088/1361-6463/ac28fa} {\bibfield  {journal} {\bibinfo  {journal} {J.
  Phys. D: App. Phys.}\ }\textbf {\bibinfo {volume} {55}},\ \bibinfo {pages}
  {103002} (\bibinfo {year} {2021})}\BibitemShut {NoStop}%
\bibitem [{\citenamefont {Bai}\ \emph {et~al.}(2024)\citenamefont {Bai},
  \citenamefont {Feng}, \citenamefont {Liu}, \citenamefont {Šmejkal},
  \citenamefont {Mokrousov},\ and\ \citenamefont {Yao}}]{yao24-review}%
  \BibitemOpen
  \bibfield  {author} {\bibinfo {author} {\bibfnamefont {L.}~\bibnamefont
  {Bai}}, \bibinfo {author} {\bibfnamefont {W.}~\bibnamefont {Feng}}, \bibinfo
  {author} {\bibfnamefont {S.}~\bibnamefont {Liu}}, \bibinfo {author}
  {\bibfnamefont {L.}~\bibnamefont {Šmejkal}}, \bibinfo {author}
  {\bibfnamefont {Y.}~\bibnamefont {Mokrousov}}, \ and\ \bibinfo {author}
  {\bibfnamefont {Y.}~\bibnamefont {Yao}},\ }\href {\doibase
  https://doi.org/10.1002/adfm.202409327} {\bibfield  {journal} {\bibinfo
  {journal} {Adv. Funct. Mater.}\ }\textbf {\bibinfo {volume} {34}},\ \bibinfo
  {pages} {2409327} (\bibinfo {year} {2024})}\BibitemShut {NoStop}%
\bibitem [{\citenamefont {{Mazin}}()}]{mazin22}%
  \BibitemOpen
  \bibfield  {author} {\bibinfo {author} {\bibfnamefont {I.~I.}\ \bibnamefont
  {{Mazin}}},\ }\href@noop {} {\ }\Eprint {http://arxiv.org/abs/2203.05000}
  {arXiv:2203.05000} \BibitemShut {NoStop}%
\bibitem [{\citenamefont {Sun}\ \emph {et~al.}(2023)\citenamefont {Sun},
  \citenamefont {Huo}, \citenamefont {Hu}, \citenamefont {Li}, \citenamefont
  {Liu}, \citenamefont {Han}, \citenamefont {Tang}, \citenamefont {Mao},
  \citenamefont {Yang}, \citenamefont {Wang}, \citenamefont {Cheng},
  \citenamefont {Yao}, \citenamefont {Zhang},\ and\ \citenamefont
  {Wang}}]{wang23-bilayer}%
  \BibitemOpen
  \bibfield  {author} {\bibinfo {author} {\bibfnamefont {H.}~\bibnamefont
  {Sun}}, \bibinfo {author} {\bibfnamefont {M.}~\bibnamefont {Huo}}, \bibinfo
  {author} {\bibfnamefont {X.}~\bibnamefont {Hu}}, \bibinfo {author}
  {\bibfnamefont {J.}~\bibnamefont {Li}}, \bibinfo {author} {\bibfnamefont
  {Z.}~\bibnamefont {Liu}}, \bibinfo {author} {\bibfnamefont {Y.}~\bibnamefont
  {Han}}, \bibinfo {author} {\bibfnamefont {L.}~\bibnamefont {Tang}}, \bibinfo
  {author} {\bibfnamefont {Z.}~\bibnamefont {Mao}}, \bibinfo {author}
  {\bibfnamefont {P.}~\bibnamefont {Yang}}, \bibinfo {author} {\bibfnamefont
  {B.}~\bibnamefont {Wang}}, \bibinfo {author} {\bibfnamefont {J.}~\bibnamefont
  {Cheng}}, \bibinfo {author} {\bibfnamefont {D.-X.}\ \bibnamefont {Yao}},
  \bibinfo {author} {\bibfnamefont {G.-M.}\ \bibnamefont {Zhang}}, \ and\
  \bibinfo {author} {\bibfnamefont {M.}~\bibnamefont {Wang}},\ }\href {\doibase
  10.1038/s41586-023-06408-7} {\bibfield  {journal} {\bibinfo  {journal}
  {Nature}\ }\textbf {\bibinfo {volume} {621}},\ \bibinfo {pages} {493}
  (\bibinfo {year} {2023})}\BibitemShut {NoStop}%
\bibitem [{\citenamefont {Li}\ \emph {et~al.}(2024)\citenamefont {Li},
  \citenamefont {Zhang}, \citenamefont {Xiang}, \citenamefont {Zhang},
  \citenamefont {Zhu},\ and\ \citenamefont {Wen}}]{wen23-trilayer}%
  \BibitemOpen
  \bibfield  {author} {\bibinfo {author} {\bibfnamefont {Q.}~\bibnamefont
  {Li}}, \bibinfo {author} {\bibfnamefont {Y.-J.}\ \bibnamefont {Zhang}},
  \bibinfo {author} {\bibfnamefont {Z.-N.}\ \bibnamefont {Xiang}}, \bibinfo
  {author} {\bibfnamefont {Y.}~\bibnamefont {Zhang}}, \bibinfo {author}
  {\bibfnamefont {X.}~\bibnamefont {Zhu}}, \ and\ \bibinfo {author}
  {\bibfnamefont {H.-H.}\ \bibnamefont {Wen}},\ }\href {\doibase
  10.1088/0256-307X/41/1/017401} {\bibfield  {journal} {\bibinfo  {journal}
  {Chinese Phys. Lett.}\ }\textbf {\bibinfo {volume} {41}},\ \bibinfo {pages}
  {017401} (\bibinfo {year} {2024})}\BibitemShut {NoStop}%
\bibitem [{\citenamefont {Pan}\ \emph {et~al.}(2021)\citenamefont {Pan},
  \citenamefont {Segedin}, \citenamefont {LaBollita}, \citenamefont {Song},
  \citenamefont {Nica}, \citenamefont {Goodge}, \citenamefont {Pierce},
  \citenamefont {Doyle}, \citenamefont {Novakov}, \citenamefont {Carrizales},
  \citenamefont {N'Diaye}, \citenamefont {Shafer}, \citenamefont {Paik},
  \citenamefont {Heron}, \citenamefont {Mason}, \citenamefont {Yacoby},
  \citenamefont {Kourkoutis}, \citenamefont {Erten}, \citenamefont {Brooks},
  \citenamefont {Botana},\ and\ \citenamefont {Mundy}}]{mundy21}%
  \BibitemOpen
  \bibfield  {author} {\bibinfo {author} {\bibfnamefont {G.~A.}\ \bibnamefont
  {Pan}}, \bibinfo {author} {\bibfnamefont {D.~F.}\ \bibnamefont {Segedin}},
  \bibinfo {author} {\bibfnamefont {H.}~\bibnamefont {LaBollita}}, \bibinfo
  {author} {\bibfnamefont {Q.}~\bibnamefont {Song}}, \bibinfo {author}
  {\bibfnamefont {E.~M.}\ \bibnamefont {Nica}}, \bibinfo {author}
  {\bibfnamefont {B.~H.}\ \bibnamefont {Goodge}}, \bibinfo {author}
  {\bibfnamefont {A.~T.}\ \bibnamefont {Pierce}}, \bibinfo {author}
  {\bibfnamefont {S.}~\bibnamefont {Doyle}}, \bibinfo {author} {\bibfnamefont
  {S.}~\bibnamefont {Novakov}}, \bibinfo {author} {\bibfnamefont {D.~C.}\
  \bibnamefont {Carrizales}}, \bibinfo {author} {\bibfnamefont {A.~T.}\
  \bibnamefont {N'Diaye}}, \bibinfo {author} {\bibfnamefont {P.}~\bibnamefont
  {Shafer}}, \bibinfo {author} {\bibfnamefont {H.}~\bibnamefont {Paik}},
  \bibinfo {author} {\bibfnamefont {J.~T.}\ \bibnamefont {Heron}}, \bibinfo
  {author} {\bibfnamefont {J.~A.}\ \bibnamefont {Mason}}, \bibinfo {author}
  {\bibfnamefont {A.}~\bibnamefont {Yacoby}}, \bibinfo {author} {\bibfnamefont
  {L.~F.}\ \bibnamefont {Kourkoutis}}, \bibinfo {author} {\bibfnamefont
  {O.}~\bibnamefont {Erten}}, \bibinfo {author} {\bibfnamefont {C.~M.}\
  \bibnamefont {Brooks}}, \bibinfo {author} {\bibfnamefont {A.~S.}\
  \bibnamefont {Botana}}, \ and\ \bibinfo {author} {\bibfnamefont {J.~A.}\
  \bibnamefont {Mundy}},\ }\href {\doibase 10.1038/s41563-021-01142-9}
  {\bibfield  {journal} {\bibinfo  {journal} {Nature Mater.}\ }\textbf
  {\bibinfo {volume} {21}},\ \bibinfo {pages} {160} (\bibinfo {year}
  {2021})}\BibitemShut {NoStop}%
\bibitem [{\citenamefont {Botana}\ \emph {et~al.}(2021)\citenamefont {Botana},
  \citenamefont {Bernardini},\ and\ \citenamefont
  {Cano}}]{review-nickelates20}%
  \BibitemOpen
  \bibfield  {author} {\bibinfo {author} {\bibfnamefont {A.~S.}\ \bibnamefont
  {Botana}}, \bibinfo {author} {\bibfnamefont {F.}~\bibnamefont {Bernardini}},
  \ and\ \bibinfo {author} {\bibfnamefont {A.}~\bibnamefont {Cano}},\ }\href
  {\doibase 10.1134/S1063776121040026} {\bibfield  {journal} {\bibinfo
  {journal} {JETP}\ }\textbf {\bibinfo {volume} {159}},\ \bibinfo {pages} {711}
  (\bibinfo {year} {2021})},\ \Eprint {http://arxiv.org/abs/2012.02764}
  {arXiv:2012.02764} \BibitemShut {NoStop}%
\bibitem [{\citenamefont {Blaha}\ \emph {et~al.}()\citenamefont {Blaha},
  \citenamefont {Schwarz}, \citenamefont {Madsen}, \citenamefont {Kvasnicka},
  \citenamefont {Luitz}, \citenamefont {Laskowski}, \citenamefont {Tran},\ and\
  \citenamefont {Marks}}]{WIEN2k}%
  \BibitemOpen
  \bibfield  {author} {\bibinfo {author} {\bibfnamefont {P.}~\bibnamefont
  {Blaha}}, \bibinfo {author} {\bibfnamefont {K.}~\bibnamefont {Schwarz}},
  \bibinfo {author} {\bibfnamefont {G.}~\bibnamefont {Madsen}}, \bibinfo
  {author} {\bibfnamefont {D.}~\bibnamefont {Kvasnicka}}, \bibinfo {author}
  {\bibfnamefont {J.}~\bibnamefont {Luitz}}, \bibinfo {author} {\bibfnamefont
  {R.}~\bibnamefont {Laskowski}}, \bibinfo {author} {\bibfnamefont
  {F.}~\bibnamefont {Tran}}, \ and\ \bibinfo {author} {\bibfnamefont {L.~D.}\
  \bibnamefont {Marks}},\ }\href@noop {} {\bibinfo  {journal} {{W}{I}{E}{N}2k,
  An Augmented Plane Wave + Local Orbitals Program for Calculating Crystal
  Properties (Karlheinz Schwarz, Techn. Universität Wien, Austria), 2018. ISBN
  3-9501031-1-2}\ }\BibitemShut {NoStop}%
\bibitem [{\citenamefont {Perdew}\ and\ \citenamefont {Zunger}(1981)}]{LDA}%
  \BibitemOpen
\bibfield  {journal} {  }\bibfield  {author} {\bibinfo {author} {\bibfnamefont
  {J.~P.}\ \bibnamefont {Perdew}}\ and\ \bibinfo {author} {\bibfnamefont
  {A.}~\bibnamefont {Zunger}},\ }\href {\doibase 10.1103/PhysRevB.23.5048}
  {\bibfield  {journal} {\bibinfo  {journal} {Phys. Rev. B}\ }\textbf {\bibinfo
  {volume} {23}},\ \bibinfo {pages} {5048} (\bibinfo {year}
  {1981})}\BibitemShut {NoStop}%
\bibitem [{\citenamefont {Perdew}\ \emph {et~al.}(1996)\citenamefont {Perdew},
  \citenamefont {Burke},\ and\ \citenamefont {Ernzerhof}}]{PBE}%
  \BibitemOpen
  \bibfield  {author} {\bibinfo {author} {\bibfnamefont {J.~P.}\ \bibnamefont
  {Perdew}}, \bibinfo {author} {\bibfnamefont {K.}~\bibnamefont {Burke}}, \
  and\ \bibinfo {author} {\bibfnamefont {M.}~\bibnamefont {Ernzerhof}},\ }\href
  {\doibase 10.1103/PhysRevLett.77.3865} {\bibfield  {journal} {\bibinfo
  {journal} {Phys. Rev. Lett.}\ }\textbf {\bibinfo {volume} {77}},\ \bibinfo
  {pages} {3865} (\bibinfo {year} {1996})}\BibitemShut {NoStop}%
\bibitem [{Note1()}]{Note1}%
  \BibitemOpen
  \bibinfo {note} {$Bmab$ is a nonconventional setting of $Cmca$ where the $c$
  axis corresponds to the long (stacking) axis. In the new ITA notation, they
  are denoted as $Bmeb$ and $Cmce$ respectively.}\BibitemShut {Stop}%
\bibitem [{\citenamefont {Rodriguez-Carvajal}\ \emph
  {et~al.}(1991)\citenamefont {Rodriguez-Carvajal}, \citenamefont
  {Fernandez-Diaz},\ and\ \citenamefont {Mart\'inez}}]{rodriguez-carvajal91}%
  \BibitemOpen
  \bibfield  {author} {\bibinfo {author} {\bibfnamefont {J.}~\bibnamefont
  {Rodriguez-Carvajal}}, \bibinfo {author} {\bibfnamefont {M.~T.}\ \bibnamefont
  {Fernandez-Diaz}}, \ and\ \bibinfo {author} {\bibfnamefont {J.~L.}\
  \bibnamefont {Mart\'inez}},\ }\href {\doibase 10.1088/0953-8984/3/19/002}
  {\bibfield  {journal} {\bibinfo  {journal} {J. Phys.: Condens. Matter}\
  }\textbf {\bibinfo {volume} {3}},\ \bibinfo {pages} {3215} (\bibinfo {year}
  {1991})}\BibitemShut {NoStop}%
\bibitem [{\citenamefont {Setyawan}\ and\ \citenamefont
  {Curtarolo}(2010)}]{Setyawan_2010}%
  \BibitemOpen
  \bibfield  {author} {\bibinfo {author} {\bibfnamefont {W.}~\bibnamefont
  {Setyawan}}\ and\ \bibinfo {author} {\bibfnamefont {S.}~\bibnamefont
  {Curtarolo}},\ }\href {\doibase
  https://doi.org/10.1016/j.commatsci.2010.05.010} {\bibfield  {journal}
  {\bibinfo  {journal} {Computational Materials Science}\ }\textbf {\bibinfo
  {volume} {49}},\ \bibinfo {pages} {299} (\bibinfo {year} {2010})}\BibitemShut
  {NoStop}%
\bibitem [{\citenamefont {Wissel}\ \emph {et~al.}(2018)\citenamefont {Wissel},
  \citenamefont {Heldt}, \citenamefont {Groszewicz}, \citenamefont {Dasgupta},
  \citenamefont {Breitzke}, \citenamefont {Donzelli}, \citenamefont {Waidha},
  \citenamefont {Fortes}, \citenamefont {Rohrer}, \citenamefont {Slater},
  \citenamefont {Buntkowsky},\ and\ \citenamefont {Clemens}}]{clemens18}%
  \BibitemOpen
  \bibfield  {author} {\bibinfo {author} {\bibfnamefont {K.}~\bibnamefont
  {Wissel}}, \bibinfo {author} {\bibfnamefont {J.}~\bibnamefont {Heldt}},
  \bibinfo {author} {\bibfnamefont {P.~B.}\ \bibnamefont {Groszewicz}},
  \bibinfo {author} {\bibfnamefont {S.}~\bibnamefont {Dasgupta}}, \bibinfo
  {author} {\bibfnamefont {H.}~\bibnamefont {Breitzke}}, \bibinfo {author}
  {\bibfnamefont {M.}~\bibnamefont {Donzelli}}, \bibinfo {author}
  {\bibfnamefont {A.~I.}\ \bibnamefont {Waidha}}, \bibinfo {author}
  {\bibfnamefont {A.~D.}\ \bibnamefont {Fortes}}, \bibinfo {author}
  {\bibfnamefont {J.}~\bibnamefont {Rohrer}}, \bibinfo {author} {\bibfnamefont
  {P.~R.}\ \bibnamefont {Slater}}, \bibinfo {author} {\bibfnamefont
  {G.}~\bibnamefont {Buntkowsky}}, \ and\ \bibinfo {author} {\bibfnamefont
  {O.}~\bibnamefont {Clemens}},\ }\href {\doibase
  10.1021/acs.inorgchem.8b00661} {\bibfield  {journal} {\bibinfo  {journal}
  {Inorg. Chem.}\ }\textbf {\bibinfo {volume} {57}},\ \bibinfo {pages} {6549}
  (\bibinfo {year} {2018})}\BibitemShut {NoStop}%
\bibitem [{\citenamefont {Wissel}\ \emph {et~al.}(2020)\citenamefont {Wissel},
  \citenamefont {Malik}, \citenamefont {Vasala}, \citenamefont {Plana-Ruiz},
  \citenamefont {Kolb}, \citenamefont {Slater}, \citenamefont {da~Silva},
  \citenamefont {Alff}, \citenamefont {Rohrer},\ and\ \citenamefont
  {Clemens}}]{clemens20}%
  \BibitemOpen
  \bibfield  {author} {\bibinfo {author} {\bibfnamefont {K.}~\bibnamefont
  {Wissel}}, \bibinfo {author} {\bibfnamefont {A.~M.}\ \bibnamefont {Malik}},
  \bibinfo {author} {\bibfnamefont {S.}~\bibnamefont {Vasala}}, \bibinfo
  {author} {\bibfnamefont {S.}~\bibnamefont {Plana-Ruiz}}, \bibinfo {author}
  {\bibfnamefont {U.}~\bibnamefont {Kolb}}, \bibinfo {author} {\bibfnamefont
  {P.~R.}\ \bibnamefont {Slater}}, \bibinfo {author} {\bibfnamefont
  {I.}~\bibnamefont {da~Silva}}, \bibinfo {author} {\bibfnamefont
  {L.}~\bibnamefont {Alff}}, \bibinfo {author} {\bibfnamefont {J.}~\bibnamefont
  {Rohrer}}, \ and\ \bibinfo {author} {\bibfnamefont {O.}~\bibnamefont
  {Clemens}},\ }\href {\doibase 10.1021/acs.chemmater.0c00193} {\bibfield
  {journal} {\bibinfo  {journal} {Chem. Mater.}\ }\textbf {\bibinfo {volume}
  {32}},\ \bibinfo {pages} {3160} (\bibinfo {year} {2020})}\BibitemShut
  {NoStop}%
\bibitem [{\citenamefont {Bernardini}\ \emph {et~al.}(2021)\citenamefont
  {Bernardini}, \citenamefont {Demourgues},\ and\ \citenamefont
  {Cano}}]{cano21-prm}%
  \BibitemOpen
  \bibfield  {author} {\bibinfo {author} {\bibfnamefont {F.}~\bibnamefont
  {Bernardini}}, \bibinfo {author} {\bibfnamefont {A.}~\bibnamefont
  {Demourgues}}, \ and\ \bibinfo {author} {\bibfnamefont {A.}~\bibnamefont
  {Cano}},\ }\href {\doibase 10.1103/PhysRevMaterials.5.L061801} {\bibfield
  {journal} {\bibinfo  {journal} {Phys. Rev. Materials}\ }\textbf {\bibinfo
  {volume} {5}},\ \bibinfo {pages} {L061801} (\bibinfo {year}
  {2021})}\BibitemShut {NoStop}%
\bibitem [{\citenamefont {Wissel}\ \emph {et~al.}(2022)\citenamefont {Wissel},
  \citenamefont {Bernardini}, \citenamefont {Oh}, \citenamefont {Vasala},
  \citenamefont {Schoch}, \citenamefont {Blaschkowski}, \citenamefont
  {Glatzel}, \citenamefont {Bauer}, \citenamefont {Clemens},\ and\
  \citenamefont {Cano}}]{cano22-chemmat}%
  \BibitemOpen
  \bibfield  {author} {\bibinfo {author} {\bibfnamefont {K.}~\bibnamefont
  {Wissel}}, \bibinfo {author} {\bibfnamefont {F.}~\bibnamefont {Bernardini}},
  \bibinfo {author} {\bibfnamefont {H.}~\bibnamefont {Oh}}, \bibinfo {author}
  {\bibfnamefont {S.}~\bibnamefont {Vasala}}, \bibinfo {author} {\bibfnamefont
  {R.}~\bibnamefont {Schoch}}, \bibinfo {author} {\bibfnamefont
  {B.}~\bibnamefont {Blaschkowski}}, \bibinfo {author} {\bibfnamefont
  {P.}~\bibnamefont {Glatzel}}, \bibinfo {author} {\bibfnamefont
  {M.}~\bibnamefont {Bauer}}, \bibinfo {author} {\bibfnamefont
  {O.}~\bibnamefont {Clemens}}, \ and\ \bibinfo {author} {\bibfnamefont
  {A.}~\bibnamefont {Cano}},\ }\href {\doibase 10.1021/acs.chemmater.2c00726}
  {\bibfield  {journal} {\bibinfo  {journal} {Chem. Mater.}\ }\textbf {\bibinfo
  {volume} {34}},\ \bibinfo {pages} {7201} (\bibinfo {year}
  {2022})}\BibitemShut {NoStop}%
\bibitem [{\citenamefont {Ling}\ \emph {et~al.}(2000)\citenamefont {Ling},
  \citenamefont {Argyriou}, \citenamefont {Wu},\ and\ \citenamefont
  {Neumeier}}]{neumeier22}%
  \BibitemOpen
  \bibfield  {author} {\bibinfo {author} {\bibfnamefont {C.~D.}\ \bibnamefont
  {Ling}}, \bibinfo {author} {\bibfnamefont {D.~N.}\ \bibnamefont {Argyriou}},
  \bibinfo {author} {\bibfnamefont {G.}~\bibnamefont {Wu}}, \ and\ \bibinfo
  {author} {\bibfnamefont {J.}~\bibnamefont {Neumeier}},\ }\href {\doibase
  https://doi.org/10.1006/jssc.2000.8721} {\bibfield  {journal} {\bibinfo
  {journal} {J. Solid State Chem.}\ }\textbf {\bibinfo {volume} {152}},\
  \bibinfo {pages} {517} (\bibinfo {year} {2000})}\BibitemShut {NoStop}%
\bibitem [{\citenamefont {Liu}\ \emph {et~al.}()\citenamefont {Liu},
  \citenamefont {Sun}, \citenamefont {Huo}, \citenamefont {Ma}, \citenamefont
  {Ji}, \citenamefont {Yi}, \citenamefont {Li}, \citenamefont {Liu},
  \citenamefont {Yu}, \citenamefont {Zhang} \emph {et~al.}}]{liu22}%
  \BibitemOpen
  \bibfield  {author} {\bibinfo {author} {\bibfnamefont {Z.}~\bibnamefont
  {Liu}}, \bibinfo {author} {\bibfnamefont {H.}~\bibnamefont {Sun}}, \bibinfo
  {author} {\bibfnamefont {M.}~\bibnamefont {Huo}}, \bibinfo {author}
  {\bibfnamefont {X.}~\bibnamefont {Ma}}, \bibinfo {author} {\bibfnamefont
  {Y.}~\bibnamefont {Ji}}, \bibinfo {author} {\bibfnamefont {E.}~\bibnamefont
  {Yi}}, \bibinfo {author} {\bibfnamefont {L.}~\bibnamefont {Li}}, \bibinfo
  {author} {\bibfnamefont {H.}~\bibnamefont {Liu}}, \bibinfo {author}
  {\bibfnamefont {J.}~\bibnamefont {Yu}}, \bibinfo {author} {\bibfnamefont
  {Z.}~\bibnamefont {Zhang}},  \emph {et~al.},\ }\href@noop {} {\bibinfo
  {journal} {arXiv:2205.00950}\ }\BibitemShut {NoStop}%
\bibitem [{\citenamefont {{LaBollita}}\ \emph {et~al.}()\citenamefont
  {{LaBollita}}, \citenamefont {{Pardo}}, \citenamefont {{Norman}},\ and\
  \citenamefont {{Botana}}}]{botana23}%
  \BibitemOpen
\bibfield  {journal} {  }\bibfield  {author} {\bibinfo {author} {\bibfnamefont
  {H.}~\bibnamefont {{LaBollita}}}, \bibinfo {author} {\bibfnamefont
  {V.}~\bibnamefont {{Pardo}}}, \bibinfo {author} {\bibfnamefont {M.~R.}\
  \bibnamefont {{Norman}}}, \ and\ \bibinfo {author} {\bibfnamefont {A.~S.}\
  \bibnamefont {{Botana}}},\ }\href@noop {} {\ }\Eprint
  {http://arxiv.org/abs/2309.17279} {arXiv:2309.17279} \BibitemShut {NoStop}%
\bibitem [{\citenamefont {Alonso}\ \emph {et~al.}(2001)\citenamefont {Alonso},
  \citenamefont {Mart\'{\i}nez-Lope}, \citenamefont {Casais}, \citenamefont
  {Garc\'{\i}a-Mu\~noz}, \citenamefont {Fern\'andez-D\'{\i}az},\ and\
  \citenamefont {Aranda}}]{alonso01}%
  \BibitemOpen
  \bibfield  {author} {\bibinfo {author} {\bibfnamefont {J.~A.}\ \bibnamefont
  {Alonso}}, \bibinfo {author} {\bibfnamefont {M.~J.}\ \bibnamefont
  {Mart\'{\i}nez-Lope}}, \bibinfo {author} {\bibfnamefont {M.~T.}\ \bibnamefont
  {Casais}}, \bibinfo {author} {\bibfnamefont {J.~L.}\ \bibnamefont
  {Garc\'{\i}a-Mu\~noz}}, \bibinfo {author} {\bibfnamefont {M.~T.}\
  \bibnamefont {Fern\'andez-D\'{\i}az}}, \ and\ \bibinfo {author}
  {\bibfnamefont {M.~A.~G.}\ \bibnamefont {Aranda}},\ }\href {\doibase
  10.1103/PhysRevB.64.094102} {\bibfield  {journal} {\bibinfo  {journal} {Phys.
  Rev. B}\ }\textbf {\bibinfo {volume} {64}},\ \bibinfo {pages} {094102}
  (\bibinfo {year} {2001})}\BibitemShut {NoStop}%
\bibitem [{\citenamefont {Bernardini}\ \emph {et~al.}(2022)\citenamefont
  {Bernardini}, \citenamefont {Bosin},\ and\ \citenamefont
  {Cano}}]{cano22-prm}%
  \BibitemOpen
  \bibfield  {author} {\bibinfo {author} {\bibfnamefont {F.}~\bibnamefont
  {Bernardini}}, \bibinfo {author} {\bibfnamefont {A.}~\bibnamefont {Bosin}}, \
  and\ \bibinfo {author} {\bibfnamefont {A.}~\bibnamefont {Cano}},\ }\href
  {\doibase 10.1103/PhysRevMaterials.6.044807} {\bibfield  {journal} {\bibinfo
  {journal} {Phys. Rev. Mater.}\ }\textbf {\bibinfo {volume} {6}},\ \bibinfo
  {pages} {044807} (\bibinfo {year} {2022})}\BibitemShut {NoStop}%
\bibitem [{\citenamefont {Xia}\ \emph {et~al.}(2022)\citenamefont {Xia},
  \citenamefont {Wu}, \citenamefont {Chen},\ and\ \citenamefont
  {Chen}}]{chen22-prb}%
  \BibitemOpen
  \bibfield  {author} {\bibinfo {author} {\bibfnamefont {C.}~\bibnamefont
  {Xia}}, \bibinfo {author} {\bibfnamefont {J.}~\bibnamefont {Wu}}, \bibinfo
  {author} {\bibfnamefont {Y.}~\bibnamefont {Chen}}, \ and\ \bibinfo {author}
  {\bibfnamefont {H.}~\bibnamefont {Chen}},\ }\href {\doibase
  10.1103/PhysRevB.105.115134} {\bibfield  {journal} {\bibinfo  {journal}
  {Phys. Rev. B}\ }\textbf {\bibinfo {volume} {105}},\ \bibinfo {pages}
  {115134} (\bibinfo {year} {2022})}\BibitemShut {NoStop}%
\bibitem [{\citenamefont {Ishiwata}\ \emph {et~al.}(2002)\citenamefont
  {Ishiwata}, \citenamefont {Azuma}, \citenamefont {Takano}, \citenamefont
  {Nishibori}, \citenamefont {Takata}, \citenamefont {Sakata},\ and\
  \citenamefont {Kato}}]{kato02}%
  \BibitemOpen
  \bibfield  {author} {\bibinfo {author} {\bibfnamefont {S.}~\bibnamefont
  {Ishiwata}}, \bibinfo {author} {\bibfnamefont {M.}~\bibnamefont {Azuma}},
  \bibinfo {author} {\bibfnamefont {M.}~\bibnamefont {Takano}}, \bibinfo
  {author} {\bibfnamefont {E.}~\bibnamefont {Nishibori}}, \bibinfo {author}
  {\bibfnamefont {M.}~\bibnamefont {Takata}}, \bibinfo {author} {\bibfnamefont
  {M.}~\bibnamefont {Sakata}}, \ and\ \bibinfo {author} {\bibfnamefont
  {K.}~\bibnamefont {Kato}},\ }\href {\doibase 10.1039/B206022A} {\bibfield
  {journal} {\bibinfo  {journal} {J. Mater. Chem.}\ }\textbf {\bibinfo {volume}
  {12}},\ \bibinfo {pages} {3733} (\bibinfo {year} {2002})}\BibitemShut
  {NoStop}%
\bibitem [{\citenamefont {Carlsson}\ \emph {et~al.}(2008)\citenamefont
  {Carlsson}, \citenamefont {Azuma}, \citenamefont {Shimakawa}, \citenamefont
  {Takano}, \citenamefont {Hewat},\ and\ \citenamefont
  {Attfield}}]{attfield08}%
  \BibitemOpen
  \bibfield  {author} {\bibinfo {author} {\bibfnamefont {S.~J.}\ \bibnamefont
  {Carlsson}}, \bibinfo {author} {\bibfnamefont {M.}~\bibnamefont {Azuma}},
  \bibinfo {author} {\bibfnamefont {Y.}~\bibnamefont {Shimakawa}}, \bibinfo
  {author} {\bibfnamefont {M.}~\bibnamefont {Takano}}, \bibinfo {author}
  {\bibfnamefont {A.}~\bibnamefont {Hewat}}, \ and\ \bibinfo {author}
  {\bibfnamefont {J.~P.}\ \bibnamefont {Attfield}},\ }\href {\doibase
  https://doi.org/10.1016/j.jssc.2007.12.037} {\bibfield  {journal} {\bibinfo
  {journal} {J. Solid State Chem.}\ }\textbf {\bibinfo {volume} {181}},\
  \bibinfo {pages} {611} (\bibinfo {year} {2008})}\BibitemShut {NoStop}%
\bibitem [{\citenamefont {Azuma}\ \emph {et~al.}(2007)\citenamefont {Azuma},
  \citenamefont {Carlsson}, \citenamefont {Rodgers}, \citenamefont {Tucker},
  \citenamefont {Tsujimoto}, \citenamefont {Ishiwata}, \citenamefont {Isoda},
  \citenamefont {Shimakawa}, \citenamefont {Takano},\ and\ \citenamefont
  {Attfield}}]{attfield07}%
  \BibitemOpen
  \bibfield  {author} {\bibinfo {author} {\bibfnamefont {M.}~\bibnamefont
  {Azuma}}, \bibinfo {author} {\bibfnamefont {S.}~\bibnamefont {Carlsson}},
  \bibinfo {author} {\bibfnamefont {J.}~\bibnamefont {Rodgers}}, \bibinfo
  {author} {\bibfnamefont {M.~G.}\ \bibnamefont {Tucker}}, \bibinfo {author}
  {\bibfnamefont {M.}~\bibnamefont {Tsujimoto}}, \bibinfo {author}
  {\bibfnamefont {S.}~\bibnamefont {Ishiwata}}, \bibinfo {author}
  {\bibfnamefont {S.}~\bibnamefont {Isoda}}, \bibinfo {author} {\bibfnamefont
  {Y.}~\bibnamefont {Shimakawa}}, \bibinfo {author} {\bibfnamefont
  {M.}~\bibnamefont {Takano}}, \ and\ \bibinfo {author} {\bibfnamefont {J.~P.}\
  \bibnamefont {Attfield}},\ }\href {\doibase 10.1021/ja074880u} {\bibfield
  {journal} {\bibinfo  {journal} {J. Am. Chem. Soc.}\ }\textbf {\bibinfo
  {volume} {129}},\ \bibinfo {pages} {14433} (\bibinfo {year}
  {2007})}\BibitemShut {NoStop}%
\bibitem [{\citenamefont {Inaguma}\ \emph {et~al.}(2011)\citenamefont
  {Inaguma}, \citenamefont {Tanaka}, \citenamefont {Tsuchiya}, \citenamefont
  {Mori}, \citenamefont {Katsumata}, \citenamefont {Ohba}, \citenamefont
  {Hiraki}, \citenamefont {Takahashi},\ and\ \citenamefont
  {Saitoh}}]{saitoh11}%
  \BibitemOpen
  \bibfield  {author} {\bibinfo {author} {\bibfnamefont {Y.}~\bibnamefont
  {Inaguma}}, \bibinfo {author} {\bibfnamefont {K.}~\bibnamefont {Tanaka}},
  \bibinfo {author} {\bibfnamefont {T.}~\bibnamefont {Tsuchiya}}, \bibinfo
  {author} {\bibfnamefont {D.}~\bibnamefont {Mori}}, \bibinfo {author}
  {\bibfnamefont {T.}~\bibnamefont {Katsumata}}, \bibinfo {author}
  {\bibfnamefont {T.}~\bibnamefont {Ohba}}, \bibinfo {author} {\bibfnamefont
  {K.-i.}\ \bibnamefont {Hiraki}}, \bibinfo {author} {\bibfnamefont
  {T.}~\bibnamefont {Takahashi}}, \ and\ \bibinfo {author} {\bibfnamefont
  {H.}~\bibnamefont {Saitoh}},\ }\href {\doibase 10.1021/ja206247j} {\bibfield
  {journal} {\bibinfo  {journal} {Journal of the American Chemical Society}\
  }\textbf {\bibinfo {volume} {133}},\ \bibinfo {pages} {16920} (\bibinfo
  {year} {2011})}\BibitemShut {NoStop}%
\bibitem [{\citenamefont {Hao}\ \emph {et~al.}(2012)\citenamefont {Hao},
  \citenamefont {Stroppa}, \citenamefont {Picozzi}, \citenamefont
  {Filippetti},\ and\ \citenamefont {Franchini}}]{franchini12-prb}%
  \BibitemOpen
  \bibfield  {author} {\bibinfo {author} {\bibfnamefont {X.~F.}\ \bibnamefont
  {Hao}}, \bibinfo {author} {\bibfnamefont {A.}~\bibnamefont {Stroppa}},
  \bibinfo {author} {\bibfnamefont {S.}~\bibnamefont {Picozzi}}, \bibinfo
  {author} {\bibfnamefont {A.}~\bibnamefont {Filippetti}}, \ and\ \bibinfo
  {author} {\bibfnamefont {C.}~\bibnamefont {Franchini}},\ }\href {\doibase
  10.1103/PhysRevB.86.014116} {\bibfield  {journal} {\bibinfo  {journal} {Phys.
  Rev. B}\ }\textbf {\bibinfo {volume} {86}},\ \bibinfo {pages} {014116}
  (\bibinfo {year} {2012})}\BibitemShut {NoStop}%
\bibitem [{\citenamefont {Hao}\ \emph {et~al.}(2014)\citenamefont {Hao},
  \citenamefont {Stroppa}, \citenamefont {Barone}, \citenamefont {Filippetti},
  \citenamefont {Franchini},\ and\ \citenamefont {Picozzi}}]{picozzi14}%
  \BibitemOpen
  \bibfield  {author} {\bibinfo {author} {\bibfnamefont {X.~F.}\ \bibnamefont
  {Hao}}, \bibinfo {author} {\bibfnamefont {A.}~\bibnamefont {Stroppa}},
  \bibinfo {author} {\bibfnamefont {P.}~\bibnamefont {Barone}}, \bibinfo
  {author} {\bibfnamefont {A.}~\bibnamefont {Filippetti}}, \bibinfo {author}
  {\bibfnamefont {C.}~\bibnamefont {Franchini}}, \ and\ \bibinfo {author}
  {\bibfnamefont {S.}~\bibnamefont {Picozzi}},\ }\href {\doibase
  10.1088/1367-2630/16/1/015030} {\bibfield  {journal} {\bibinfo  {journal}
  {New J. Phys.}\ }\textbf {\bibinfo {volume} {16}},\ \bibinfo {pages} {015030}
  (\bibinfo {year} {2014})}\BibitemShut {NoStop}%
\bibitem [{\citenamefont {{Yuan}}\ \emph {et~al.}()\citenamefont {{Yuan}},
  \citenamefont {{Zhang}}, \citenamefont {{Mera}},\ and\ \citenamefont
  {{Zunger}}}]{zunger22}%
  \BibitemOpen
  \bibfield  {author} {\bibinfo {author} {\bibfnamefont {L.-D.}\ \bibnamefont
  {{Yuan}}}, \bibinfo {author} {\bibfnamefont {X.}~\bibnamefont {{Zhang}}},
  \bibinfo {author} {\bibfnamefont {C.}~\bibnamefont {{Mera}}}, \ and\ \bibinfo
  {author} {\bibfnamefont {A.}~\bibnamefont {{Zunger}}},\ }\href@noop {} {\
  }\Eprint {http://arxiv.org/abs/2211.09921} {arXiv:2211.09921} \BibitemShut
  {NoStop}%
\bibitem [{\citenamefont {Kamegashira}\ \emph {et~al.}(2006)\citenamefont
  {Kamegashira}, \citenamefont {Satoh}, \citenamefont {Ito},\ and\
  \citenamefont {Shimono}}]{shimono06}%
  \BibitemOpen
  \bibfield  {author} {\bibinfo {author} {\bibfnamefont {N.}~\bibnamefont
  {Kamegashira}}, \bibinfo {author} {\bibfnamefont {H.}~\bibnamefont {Satoh}},
  \bibinfo {author} {\bibfnamefont {Y.}~\bibnamefont {Ito}}, \ and\ \bibinfo
  {author} {\bibfnamefont {A.}~\bibnamefont {Shimono}},\ }\href {\doibase
  https://doi.org/10.1016/j.jallcom.2004.12.059} {\bibfield  {journal}
  {\bibinfo  {journal} {J. Alloys Compds}\ }\textbf {\bibinfo {volume}
  {408-412}},\ \bibinfo {pages} {589} (\bibinfo {year} {2006})}\BibitemShut
  {NoStop}%
\bibitem [{\citenamefont {Yamada}\ \emph {et~al.}(1989)\citenamefont {Yamada},
  \citenamefont {Matsuda}, \citenamefont {Endoh}, \citenamefont {Keimer},
  \citenamefont {Birgeneau}, \citenamefont {Onodera}, \citenamefont {Mizusaki},
  \citenamefont {Matsuura},\ and\ \citenamefont {Shirane}}]{shirane89}%
  \BibitemOpen
  \bibfield  {author} {\bibinfo {author} {\bibfnamefont {K.}~\bibnamefont
  {Yamada}}, \bibinfo {author} {\bibfnamefont {M.}~\bibnamefont {Matsuda}},
  \bibinfo {author} {\bibfnamefont {Y.}~\bibnamefont {Endoh}}, \bibinfo
  {author} {\bibfnamefont {B.}~\bibnamefont {Keimer}}, \bibinfo {author}
  {\bibfnamefont {R.~J.}\ \bibnamefont {Birgeneau}}, \bibinfo {author}
  {\bibfnamefont {S.}~\bibnamefont {Onodera}}, \bibinfo {author} {\bibfnamefont
  {J.}~\bibnamefont {Mizusaki}}, \bibinfo {author} {\bibfnamefont
  {T.}~\bibnamefont {Matsuura}}, \ and\ \bibinfo {author} {\bibfnamefont
  {G.}~\bibnamefont {Shirane}},\ }\href {\doibase 10.1103/PhysRevB.39.2336}
  {\bibfield  {journal} {\bibinfo  {journal} {Phys. Rev. B}\ }\textbf {\bibinfo
  {volume} {39}},\ \bibinfo {pages} {2336} (\bibinfo {year}
  {1989})}\BibitemShut {NoStop}%
\bibitem [{\citenamefont {{Romero de Paz}}\ \emph
  {et~al.}(1999{\natexlab{a}})\citenamefont {{Romero de Paz}}, \citenamefont
  {Fernández-Díaz}, \citenamefont {{Hernández Velasco}}, \citenamefont
  {{Sáez Puche}},\ and\ \citenamefont {Martínez}}]{martinez1}%
  \BibitemOpen
  \bibfield  {author} {\bibinfo {author} {\bibfnamefont {J.}~\bibnamefont
  {{Romero de Paz}}}, \bibinfo {author} {\bibfnamefont {M.}~\bibnamefont
  {Fernández-Díaz}}, \bibinfo {author} {\bibfnamefont {J.}~\bibnamefont
  {{Hernández Velasco}}}, \bibinfo {author} {\bibfnamefont {R.}~\bibnamefont
  {{Sáez Puche}}}, \ and\ \bibinfo {author} {\bibfnamefont {J.}~\bibnamefont
  {Martínez}},\ }\href {\doibase 10.1006/jssc.1998.7973} {\bibfield  {journal}
  {\bibinfo  {journal} {J. Solid State Chem.}\ }\textbf {\bibinfo {volume}
  {142}},\ \bibinfo {pages} {29} (\bibinfo {year}
  {1999}{\natexlab{a}})}\BibitemShut {NoStop}%
\bibitem [{\citenamefont {{Romero de Paz}}\ \emph
  {et~al.}(1999{\natexlab{b}})\citenamefont {{Romero de Paz}}, \citenamefont
  {{Hernández Velasco}}, \citenamefont {Fernández-Díaz}, \citenamefont
  {Porcher}, \citenamefont {Martínez},\ and\ \citenamefont {{Sáez
  Puche}}}]{martinez2}%
  \BibitemOpen
  \bibfield  {author} {\bibinfo {author} {\bibfnamefont {J.}~\bibnamefont
  {{Romero de Paz}}}, \bibinfo {author} {\bibfnamefont {J.}~\bibnamefont
  {{Hernández Velasco}}}, \bibinfo {author} {\bibfnamefont {M.}~\bibnamefont
  {Fernández-Díaz}}, \bibinfo {author} {\bibfnamefont {P.}~\bibnamefont
  {Porcher}}, \bibinfo {author} {\bibfnamefont {J.}~\bibnamefont {Martínez}},
  \ and\ \bibinfo {author} {\bibfnamefont {R.}~\bibnamefont {{Sáez Puche}}},\
  }\href {\doibase 10.1006/jssc.1999.8461} {\bibfield  {journal} {\bibinfo
  {journal} {J. Solid State Chem.}\ }\textbf {\bibinfo {volume} {148}},\
  \bibinfo {pages} {361} (\bibinfo {year} {1999}{\natexlab{b}})}\BibitemShut
  {NoStop}%
\bibitem [{\citenamefont {{Romero de Paz}}\ \emph {et~al.}(2000)\citenamefont
  {{Romero de Paz}}, \citenamefont {Martínez},\ and\ \citenamefont {{Sáez
  Puche}}}]{martinez3}%
  \BibitemOpen
  \bibfield  {author} {\bibinfo {author} {\bibfnamefont {J.}~\bibnamefont
  {{Romero de Paz}}}, \bibinfo {author} {\bibfnamefont {J.}~\bibnamefont
  {Martínez}}, \ and\ \bibinfo {author} {\bibfnamefont {R.}~\bibnamefont
  {{Sáez Puche}}},\ }\href {\doibase 10.1016/S0925-8388(00)00588-0} {\bibfield
   {journal} {\bibinfo  {journal} {J. Alloys Compds}\ }\textbf {\bibinfo
  {volume} {303-304}},\ \bibinfo {pages} {293} (\bibinfo {year}
  {2000})}\BibitemShut {NoStop}%
\bibitem [{\citenamefont {Sosnowska}\ \emph {et~al.}(1982)\citenamefont
  {Sosnowska}, \citenamefont {Neumaier},\ and\ \citenamefont
  {Steichele}}]{steichele82}%
  \BibitemOpen
  \bibfield  {author} {\bibinfo {author} {\bibfnamefont {I.}~\bibnamefont
  {Sosnowska}}, \bibinfo {author} {\bibfnamefont {T.~P.}\ \bibnamefont
  {Neumaier}}, \ and\ \bibinfo {author} {\bibfnamefont {E.}~\bibnamefont
  {Steichele}},\ }\href {\doibase 10.1088/0022-3719/15/23/020} {\bibfield
  {journal} {\bibinfo  {journal} {J. Phys. C: Solid State Phys.}\ }\textbf
  {\bibinfo {volume} {15}},\ \bibinfo {pages} {4835} (\bibinfo {year}
  {1982})}\BibitemShut {NoStop}%
\bibitem [{\citenamefont {Burns}\ \emph {et~al.}(2020)\citenamefont {Burns},
  \citenamefont {Paull}, \citenamefont {Juraszek}, \citenamefont {Nagarajan},\
  and\ \citenamefont {Sando}}]{sando20}%
  \BibitemOpen
  \bibfield  {author} {\bibinfo {author} {\bibfnamefont {S.~R.}\ \bibnamefont
  {Burns}}, \bibinfo {author} {\bibfnamefont {O.}~\bibnamefont {Paull}},
  \bibinfo {author} {\bibfnamefont {J.}~\bibnamefont {Juraszek}}, \bibinfo
  {author} {\bibfnamefont {V.}~\bibnamefont {Nagarajan}}, \ and\ \bibinfo
  {author} {\bibfnamefont {D.}~\bibnamefont {Sando}},\ }\href {\doibase
  https://doi.org/10.1002/adma.202003711} {\bibfield  {journal} {\bibinfo
  {journal} {Advanced Materials}\ }\textbf {\bibinfo {volume} {32}},\ \bibinfo
  {pages} {2003711} (\bibinfo {year} {2020})}\BibitemShut {NoStop}%
\bibitem [{\citenamefont {Milivojević}\ \emph {et~al.}(2024)\citenamefont
  {Milivojević}, \citenamefont {Orozović}, \citenamefont {Picozzi},
  \citenamefont {Gmitra},\ and\ \citenamefont {Stavrić}}]{Milivojević_2024}%
  \BibitemOpen
  \bibfield  {author} {\bibinfo {author} {\bibfnamefont {M.}~\bibnamefont
  {Milivojević}}, \bibinfo {author} {\bibfnamefont {M.}~\bibnamefont
  {Orozović}}, \bibinfo {author} {\bibfnamefont {S.}~\bibnamefont {Picozzi}},
  \bibinfo {author} {\bibfnamefont {M.}~\bibnamefont {Gmitra}}, \ and\ \bibinfo
  {author} {\bibfnamefont {S.}~\bibnamefont {Stavrić}},\ }\href {\doibase
  10.1088/2053-1583/ad4c73} {\bibfield  {journal} {\bibinfo  {journal} {2D
  Materials}\ }\textbf {\bibinfo {volume} {11}},\ \bibinfo {pages} {035025}
  (\bibinfo {year} {2024})}\BibitemShut {NoStop}%
\bibitem [{\citenamefont {Jo}\ \emph {et~al.}()\citenamefont {Jo},
  \citenamefont {Go}, \citenamefont {Mokrousov}, \citenamefont {Oppeneer},
  \citenamefont {Cheong},\ and\ \citenamefont {Lee}}]{lee24}%
  \BibitemOpen
  \bibfield  {author} {\bibinfo {author} {\bibfnamefont {D.}~\bibnamefont
  {Jo}}, \bibinfo {author} {\bibfnamefont {D.}~\bibnamefont {Go}}, \bibinfo
  {author} {\bibfnamefont {Y.}~\bibnamefont {Mokrousov}}, \bibinfo {author}
  {\bibfnamefont {P.~M.}\ \bibnamefont {Oppeneer}}, \bibinfo {author}
  {\bibfnamefont {S.-W.}\ \bibnamefont {Cheong}}, \ and\ \bibinfo {author}
  {\bibfnamefont {H.-W.}\ \bibnamefont {Lee}},\ }\href
  {https://arxiv.org/abs/2410.17386} {\ }\Eprint
  {http://arxiv.org/abs/2410.17386} {arXiv:2410.17386} \BibitemShut {NoStop}%
\end{thebibliography}%

\end{document}